\def\lsim{~\rlap{$<$}{\lower 1.0ex\hbox{$\sim$}}}

\documentclass[preprint]{aastex}
\shorttitle{Discovery of TYC 2949-00557-1b}
\shortauthors{Scott W. Fleming et al.}
\begin{document}
\title{Discovery of a Low-Mass Companion to a Metal-Rich F Star with the MARVELS Pilot Project}

\author{Scott W. Fleming\altaffilmark{1}, Jian Ge\altaffilmark{1}, Suvrath Mahadevan\altaffilmark{1,10,11}, Brian Lee\altaffilmark{1}, Jason D. Eastman\altaffilmark{2}, Robert J. Siverd\altaffilmark{2}, B. Scott Gaudi\altaffilmark{2}, Andrzej Niedzielski\altaffilmark{3}, Thirupathi Sivarani\altaffilmark{4}, Keivan Stassun\altaffilmark{5,6}, Alex Wolszczan\altaffilmark{10,11}, Rory Barnes\altaffilmark{7}, Bruce Gary\altaffilmark{5}, Duy Cuong Nguyen\altaffilmark{1}, Robert C. Morehead\altaffilmark{1}, Xiaoke Wan\altaffilmark{1}, Bo Zhao\altaffilmark{1}, Jian Liu\altaffilmark{1}, Pengcheng Guo\altaffilmark{1}, Stephen R. Kane\altaffilmark{1,8}, Julian C. van Eyken\altaffilmark{1,8}, Nathan M. De Lee\altaffilmark{1}, Justin R. Crepp\altaffilmark{1,9}, Alaina C. Shelden\altaffilmark{1,12}, Chris Laws\altaffilmark{7}, John P. Wisniewski\altaffilmark{7}, Donald P. Schneider\altaffilmark{10,11}, Joshua Pepper\altaffilmark{5}, Stephanie A. Snedden\altaffilmark{12}, Kaike Pan\altaffilmark{12}, Dmitry Bizyaev\altaffilmark{12}, Howard Brewington\altaffilmark{12}, Olena Malanushenko\altaffilmark{12}, Viktor Malanushenko\altaffilmark{12}, Daniel Oravetz\altaffilmark{12}, Audrey Simmons\altaffilmark{12}, Shannon Watters\altaffilmark{12,13}}

\email{scfleming@astro.ufl.edu}
\altaffiltext{1}{Dept. of Astronomy, University of Florida, 211 Bryant Space Science Center, Gainesville, FL, 326711-2055 USA}
\altaffiltext{2}{Department of Astronomy, The Ohio State University, 140 West 18th Avenue, Columbus, OH 43210}
\altaffiltext{3}{Toru\'{n} Center for Astronomy, Nicolaus Copernicus University, ul. Gagarina 11, 87-100, Toru\'{n}, Poland}
\altaffiltext{4}{Indian Institute of Astrophysics, Bangalore 560034, India}
\altaffiltext{5}{Department of Physics and Astronomy, Vanderbilt University, Nashville, TN 37235}
\altaffiltext{6}{Fisk University, Department of Physics, 1000 17th Ave. N., Nashville, TN 37208.}
\altaffiltext{7}{Department of Astronomy, University of Washington, Box 351580, Seattle, WA 98195, USA}
\altaffiltext{8}{NASA Exoplanet Science Institute, Caltech, MS 100-22, 770 South Wilson Avenue, Pasadena, CA 91125, USA}
\altaffiltext{9}{Department of Astronomy, California Institute of Technology, 1200 E. California Blvd., Pasadena, CA 91125, USA}
\altaffiltext{10}{Department of Astronomy and Astrophysics, The Pennsylvania State University, 525 Davey Laboratory, University Park, PA 16802, USA.}
\altaffiltext{11}{Center for Exoplanets and Habitable Worlds, The Pennsylvania State University, University Park, PA 16802, USA.}
\altaffiltext{12}{Apache Point Observatory, P.O. Box 59, Sunspot, NM 88349-0059}
\altaffiltext{13}{Institute for Astronomy, 34 Ohia Ku St., Pukalani, HI 96768-8288, USA}

\begin{abstract}
We report the discovery of a low-mass companion orbiting the
metal-rich, main sequence F star TYC 2949-00557-1 during the MARVELS
(Multi-object APO Radial Velocity Exoplanet Large-area Survey) Pilot Project.  The host star has an
effective temperature $T_{\rm eff}=6135 \pm 40~{\rm K}$, ${\rm
log}g=4.4 \pm 0.1$ and [Fe/H]=$0.32 \pm 0.01$, indicating a mass of
$M=1.25 \pm 0.09~M_\odot$ and $R=1.15 \pm 0.15~R_\odot$.  The
companion has an orbital period of $5.69449 \pm 0.00023$ days and
straddles the hydrogen burning limit with a minimum mass of 64 $M_J$,
and may thus be an example of the rare class of brown dwarfs orbiting
at distances comparable to those of ``Hot Jupiters.''  We present
relative photometry that demonstrates the host star is photometrically
stable at the few millimagnitude level on time scales of hours
to years, and rules out transits for a companion of radius $\ga 0.8~R_J$ at the $95\%$ confidence level.  Tidal analysis of the system suggests that the star and
companion are likely in a double synchronous state where both
rotational and orbital synchronization have been achieved.  This is
the first low-mass companion detected with a multi-object, dispersed,
fixed-delay interferometer.
\end{abstract}

\section{Introduction}
Studies of the frequency, parameter distributions and correlations of
extrasolar planets require homogeneous samples of hundreds of planets
to obtain statistically significant results.  Moreover, such a sample
must have well-understood completeness limits, selection effects and
biases, which are easiest to obtain from a single, large-scale survey.
Given current constraints on the frequency of giant planets, detection
of such a large sample of planetary systems generally requires a
precision radial velocity (RV) survey of many thousands of stars.
Such a survey also provides a wealth of ancillary science.  In
particular, it is exquisitely sensitive to more massive companions,
and because it targets a large and broad sample of host stars, it is
naturally sensitive to rare binary systems in poorly explored regions
of parameter space.

Of particular interest are constraints on the frequency and parameter
distributions of low-mass companions to solar-type stars with masses
near the hydrogen burning limit.  One of the early results from
precise RV searches was the apparent paucity of brown dwarf companions
with minimum masses ($12~M_{J} \la m\sin i \la 80~M_{J}$) at
separations of $a\la 5$~AU, relative to more massive stellar
companions and less massive planetary companions \citep{mar2000}.
Note that we denote $i$ as the inclination angle between the
companion's orbital angular momentum vector and the line-of-sight, and we reserve $I$ as the 
inclination angle of the stellar rotation axis to the line-of-sight.  While the frequency
of brown dwarf companions at larger separations is still relatively
uncertain (e.g., \citealt{met2009}), a meta-analysis of sets of known
companions to solar-type stars by \citet{gre2006}, with corrections
for observational bias, confirmed the lack of brown dwarfs at close
separations. These authors place the `driest' part of the brown dwarf
desert at $\sim 20-50~M_J$, with a frequency of companions $\la 0.5\%$
in this range of masses. 

Although there has been a steady increase in the number of known
brown dwarf candidates via the
RV technique \citep{mar2001,udr2002,end2004,patel07,wit2009,kan2009,jen2009,nie2009,omiya09}, most
of these detections have been at separations $a \gtrsim$ 0.8 AU.
Notable exceptions include the transiting brown dwarf CoRoT-Exo-3b with a
period of $\sim 4$ days orbiting an F3V star \citep{del2008}, and HD41004Bb
with a period of $\sim 1$ day orbiting the M dwarf component of a K-M
binary system \citep{san2002}.  Brown dwarfs at such
short orbital separations are of particular interest for several
reasons.  First, the frequency of such systems as a function of their
physical and orbital parameters provide diagnostics that may be able to 
distinguish between the various mechanisms that have been
invoked for their formation and dynamical evolution (e.g., \citealt{arm2002,matzner05}).  In particular, these systems
offer observational constraints on the poorly-understood theory of
tidal interactions between host stars and close companions
(e.g., \citealt{maz2008,pon2009}).  Second, these systems are much more likely to transit
than their longer-period counterparts, as the transit probability is inversely proportional to orbital separation. 
Transiting systems yield valuable measurements on the masses, radii, and mean densities of brown dwarfs \citep{sta2006,sta2007,del2008}. 

Here we report the discovery of a candidate short-period, brown
dwarf companion to the metal-rich star TYC 2949-00557-1, a main sequence F star with apparent brightness $V \sim 12.1$.  This companion
was discovered as part of the Multi-object APO Radial Velocity Exoplanet Large-area Survey (MARVELS) Pilot Project (hereafter MPP).   The MPP used the W. M. Keck Exoplanet Tracker (Keck ET)
instrument \citep{gespie2006} on the Sloan Digital Sky Survey (SDSS) 2.5m telescope \citep{gun2006} at the Apache Point Observatory.  
The Keck ET instrument is a multi-object (59 targets per exposure), dispersed fixed-delay interferometer \citep[DFDI,][]{gee2002,ge2002,ers2002,ers2003}.
In this instrument, fiber-fed starlight from the telescope is first passed through an iodine
cell that acts as a stable wavelength reference.  This 
light is then fed through a fixed-delay interferometer controlled via a piezoelectric transducer (PZT), and finally through a spectrograph that has a spectral resolution of R $ = 5100$. 
Radial velocity information is then imprinted in the phases of the 
fringes perpendicular to the dispersion axis of the spectrum due to a fixed variation in the interferometer delay along this direction.

The primary goal of the MPP was to demonstrate 
a fully-integrated DFDI instrument capable of observing
multiple stars in a single exposure.  The second goal was to
demonstrate that such an instrument is capable of achieving Doppler RV
precision sufficient for extrasolar planet detection.  The final goal
was to formulate an operations procedure for conducting a survey using the Keck ET
instrument in an efficient manner.  These three goals were necessary
preparations for the MARVELS \citep{ge2009} survey: a multi-object,
DFDI, extrasolar planet survey that is part of Sloan Digital Sky Survey III (SDSS-III).\footnote{http://www.sdss3.org/}  

The MARVELS Pilot Project was conducted in 2007, and consisted of 5-38
observations of 708 stars taken over 1-5 month baselines. Although the
primary purpose of the MPP was to lay the groundwork for the full
MARVELS survey, the cadence, number, and precision of the MPP was
nevertheless sufficient to detect massive companions to a number of
the target stars.  The RV data for each star was initially fit using the RVSIM program \citep{kan2007}.  The companion to TYC 2949-00557-1 emerged as an
excellent candidate, and here we describe the discovery radial velocity data, as well as additional spectroscopic and photometric data acquired to confirm the companion and further characterize the
host star. TYC 2949-00557-1b is the first low-mass companion detected
with a multi-object, dispersed, fixed-delay interferometer; previous
observations with a single-object DFDI instrument at Kitt Peak
National Observatory resulted in the first extrasolar planet
discovered via this technique \citep{ge2006}, as well as the first confirmed planet via DFDI \citep{van2004} and the ability to measure precise, absolute radial velocities with DFDI \citep{mah2008}.

\section{Doppler Observations}
\subsection{MPP Observations}

The MPP targeted 708 F, G, and K dwarfs with $7.6 < V < 12$ in 12
different fields, each containing 59 stars.  Data for each field was
processed simultaneously using a pipeline developed for multi-object
DFDI instruments \citep[see][for details of basic DFDI processing
steps.]{van2004,ge2006,mah2008} For each target, we determine a
``quality factor'' ($QF$), which we define as
\begin{equation}
QF = \frac{\rm{RMS}\left(X - \langle {X}
\rangle\right)}{\rm{MEDIAN}\left(\sigma_{X}\right)},
\label{eqn:qf}
\end{equation}
where $X$ represents the RV measurements and RMS is the
root-mean-square residual.  Targets that have $QF$ $>$ 25 either have
intrinsically large RV variability, or they are stars with
line-of-sight rotational velocities $\gtrsim$10 km s$^{-1}$, from
which we are unable to extract precise RV measurements.  TYC
2949-00557-1 was identified from a field with 58 other targets as
having a $QF$ of 37.20, in this case indicative of its binary nature.
A total of 14 usable Doppler RV measurements were obtained spanning
116 days from Jan-Apr 2007.  Table \ref{aporvs} contains the 
barycentric Julian date using the Barycentric Dynamical Time standard (BJD$_{\rm{TDB}}$), measured Doppler RV and
associated errors for all 14 observations.  The Doppler velocities
presented in Table \ref{aporvs} are absolute velocities calibrated to
the solar spectrum, and are the sum of the barycentric velocity of the
system and the additional Doppler variability caused by the companion.
In order to provide an indication of the true level of systematic
errors in the MPP data for objects in this field, we note that the 47
(likely constant) targets with $QF < 25$ in this field have a median
quality factor of 3.25.

\begin{figure}
\plotone{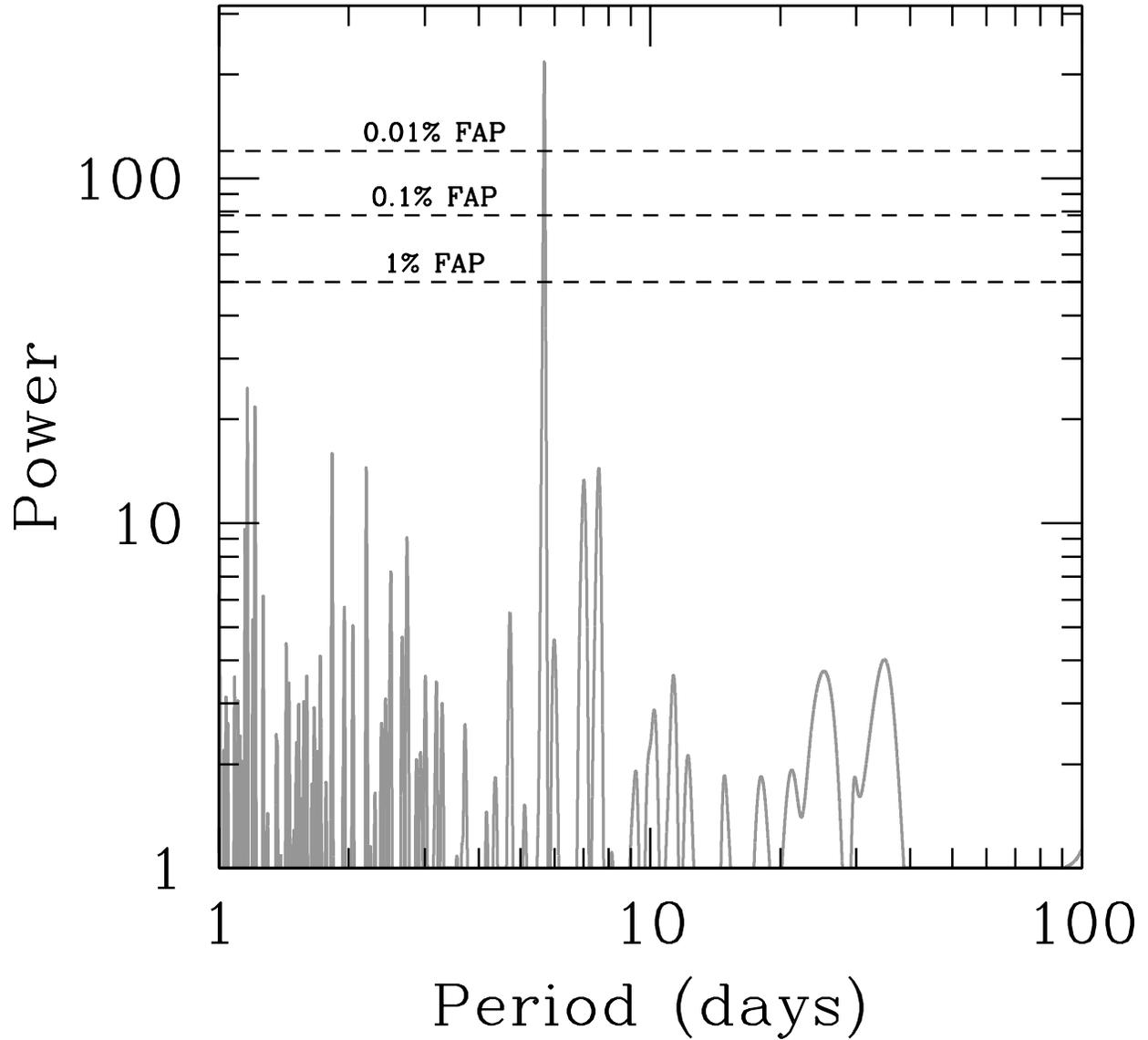}
\caption{Lomb-Scargle periodogram for the MPP data (14 epochs).  A clear and highly significant peak at $P=5.68~{\rm days}$ can be seen, with a power of $\sim 217$,
which has a false alarm probability based on scrambling the data of $<0.01\%$.  Powers corresponding to false alarm probabilities of 1\%, 0.1\%, and 0.01\% are also shown.
\label{power}}
\end{figure}

We searched for periodicity in the Doppler data for TYC 2949-00557-1
using a Lomb-Scargle periodogram with floating mean
\citep{lomb76,scargle82,cumming99}.  The resulting power spectrum is
shown in Fig.\ \ref{power}, revealing a clear peak at $P=5.68~{\rm
days}$ with a power of $\sim 217$.  To assess the significance of this
peak, we ran a Monte Carlo simulation with $10^5$ trials.  For each
trial, we scrambled the times of the data points, computed the
periodogram, and recorded the most significant peak.  We found no
trials with power greater than that of actual data, indicating a false
alarm probability of $<10^{-5}$.  

The advantage of a multi-object instrument is that the multiple targets
that are observed simultaneously can be used to check for common systematics trends
in data.  Thus, as an additional check on the reliability of the observations, we constructed
periodograms for the other 58 objects on the plate and calculated the
power for each object at the period of the suspected companion.  Any common systematics present in the data
due to the sampling rate or instrumental effects will result in significant power
at a common period for other targets.  None of the other 58
targets have significant power at the period of TYC 2949-00557-1; the next strongest candidate has an FAP at that period of 77\% and
the other targets have FAPs $>$ 99\%.

The best-fit amplitude derived from the periodogram is $\sim 5500~{\rm m~s^{-1}}$,
indicating a minimum mass in the brown dwarf regime for an FGK-type primary.  
Given that brown dwarf companions in this period range are rare, we decided to obtain additional precise radial velocity measurements, high-resolution
spectra, absolute photometry, and precise relative photometry time series, in order
to better characterize the primary and ascertain the nature of the companion.

\subsection{KPNO Doppler Observations}
Observations for the purpose of confirming the Doppler variability and orbit were conducted with the Exoplanet Tracker (ET) instrument \citep{ge2006} at the Kitt Peak National Observatory with the 2.1m telescope.  Two observations separated by several hours were taken each night over seven consecutive nights starting on Oct.\ 10, 2008.  A total of eleven usable epochs were obtained.  Integrations consisted of 60-min exposures bracketed by exposures of a Tungsten lamp passing through an iodine gas cell that acts as a calibration for instrument drift.  Each arm of the interferometer produces a DFDI spectrum from which radial velocities are measured.  The two beams are processed separately, and their measured radial velocities are combined via a weighted average based on the RV uncertainties.  Table \ref{kpnorvs} contains the dates and velocities for the KPNO ET measurements.  Unlike the results from the MPP, the velocities presented in Table \ref{kpnorvs} are relative RVs, i.e., measured relative to one of the epochs.  Note that none of the velocities are exactly zero because an instrumental drift has been subtracted off based on the calibration lamps, and they are zeroed to a different epoch than the star.  Because these data are relative RVs, an offset exists between the values in Table \ref{aporvs} and Table \ref{kpnorvs} that must be included as an additional parameter when performing a combined analysis of the two data sets.

\subsection{HET Doppler Observations}
Observations of the candidate were also conducted using the R=60,000 mode of the HRS spectrograph \citep{tull98} on the Hobby-Eberly Telescope (HET) telescope \citep{ram1998} in the queue scheduled mode \citep{she2007}.  The spectra consisted of 46 echelle orders recorded on the ``blue'' CCD detector (407.6-592 nm) and 24 orders on the ``red'' one (602-783.8 nm). The spectral data used for RV measurements were extracted from the 17 orders that cover the 505-592 nm range of the iodine spectrum.   A total of ten Doppler RV measurements were obtained spanning 83 days from Dec 2008 through Feb 2009.  The starlight was passed through an iodine cell to provide a stable reference to calibrate instrument drift.  Two exposures without the iodine cell were taken to act as stellar templates.  Due to the faintness of the target, the RVs were computed relative to each template and a mean value was determined.  The results for both templates agree to within three sigma.  Table \ref{hetrvs} contains the dates and velocities for the HET measurements.  Similar to the results from the KPNO ET in Table \ref{kpnorvs}, these velocities are relative RVs, and therefore an offset exists between these values and the ones in both Table \ref{aporvs} and Table \ref{kpnorvs}.

\subsection{Combined RV Analysis\label{combined}}

In order to check for consistency, we first fit the MPP, KPNO and HET datasets individually to a seven parameter RV fit, where the seven parameters are the velocity semi-amplitude $K$, eccentricity $e$, argument of periastron $\omega$, period $P$, time of inferior conjunction of the companion $T_c$, velocity zero point $\gamma$, and linear slope $\dot\gamma$ (in order to allow for additional companions or systematic drifts).  The best-fit solution was found using a hybrid downhill-simplex fit to the nonlinear parameters and an exact (linear) fit to the linear parameters.  There are 14 MPP points, 11 KPNO points and 10 HET points, so there are seven, four and three degrees of freedom (dof), respectively.

For the MPP fit, we find a $\chi^2/{\rm dof}$ of 44.  Given that the data points basically follow the model, and that the more precise HET data (whose error bars are overestimated, see below) fit the model well, the large $\chi^2/{\rm dof}$ indicates that there are systematic uncertainties in addition to the photon noise, and thus the errors are severely underestimated.  Given the large median $QF$=3.25 found for the majority of the (likely constant) stars in this field, this result is not surprising. Indeed, there is a known systematic error in DFDI when utilizing an iodine cell in the stellar beam path \citep{van2010}.  For the KPNO fit we find a $\chi^2/{\rm dof}$ of 6.96, once again indicating the uncertainties are underestimated.  For the HET fit, we find a $\chi^2/{\rm dof}$ of 0.09, therefore the uncertainties are likely overestimated for the HET dataset.  It is worth noting that a statistically significant slope is found when fitting the HET data.  Fitting the HET data without a slope produces a significantly worse $\chi^2/{\rm dof}$.

We performed additional RV fitting to test the significance of the HET slope.  Fitting the HET data with no slope and eccentricity forced to zero still results in a $\chi^2/{\rm dof} \ll 1$.  Since this is the simplest model of an orbiting companion, it confirms the HET uncertainties are overestimated.  Several different models, in which slope is a free parameter, eccentricity is a free parameter, or both are free parameters, all result in lower $\chi^2/{\rm dof}$.  There is no evidence of nonzero eccentricity in any of the best-fit solutions, but a significant slope is found in all cases when left as a free parameter.  To be self-consistent, we allow for slopes in the MPP and KPNO data fitting as well, and note that their best-fit slopes are consistent with the HET value, but are much more poorly constrained due to the much larger RV uncertainties in those data sets.  As a final check, we fit all data sets with eccentricity forced to zero and no slope, and find that the other orbital parameters are not qualitatively different from the result where slope is left as a free parameter.  We therefore choose the case of zero eccentricity and non-zero slope as our preferred solution.

Uncertainties in the fitting parameters will be inaccurate if it is determined using misestimated RV errors.  It is therefore important to attempt to correct the errors such that $\chi^2/{\rm dof}\sim 1$.  However, given that we do not know why the errors are misestimated, particularly in the case of underestimated errors, the appropriate method to correct the uncertainties is not clear.  Our approach was to try several different ways of correcting the errors to force $\chi^2/{\rm dof}=1$.  Specifically we investigated four different cases for treatment of the MPP (KPNO data come from a similar pipeline) and HET RV uncertainties: a scaling of the errors by a constant factor, an addition in quadrature of a constant error, a removal of suspected outliers based on the magnitude of the RV uncertainty followed by scaling, and a treatment of all data points with a constant error value.  Ten Markov Chain Monte Carlo (MCMC) simulations were run for each case.  The starting values for the parameters in these chains were chosen to span a range that is large with respect to the expected $1\sigma$ uncertainty, and the chains were stopped after reaching convergence as defined in \citet{for2006}, and then the chains were merged.

Analyzing the MPP and HET data separately yields discrepant periods at the $\sim2\sigma$ level for all cases of error treatment.  We also found that the choice of error treatment can affect the derived value of $e \cos{\omega}$ from the MPP data as well at the $\sim 1\sigma$ level.  However, the other parameters from the MPP data, as well as all parameter values from the HET data, were not significantly affected by different treatments of the uncertainties.  From this test, we conclude that there is no strong justification for removing any data points from the fit, so we conducted the final joint analysis of all three datasets where the each set of errors are scaled by a constant factor.

The MPP errors are scaled by a factor of 6.64, the KPNO errors are scaled by 2.64, and the HET errors are scaled by a factor of 0.3 such that the reduced $\chi^2$ is $\sim$ 1 when each is fit independently.  The fit including eccentricity as a free-parameter is consistent with zero eccentricity, therefore we run a second fit that forces $e = 0$.  In that case the error scalings are factors of \{6.82, 2.71, 0.24\} for MPP, KPNO and HET, respectively.  We further scale the errors of all three data sets by a factor of 1.30 for the case where eccentricity is left as a free parameter, and 1.25 for the case where eccentricity is fixed to zero.  This scaling is done so that the $\chi^2/{\rm dof}=1$ in the combined fit, and is necessary due to the systematics present in the data sets.  Given the close separation of the companion, it is expected that the orbit has been tidally circularized, consistent with our findings.  We therefore treat the case with eccentricity fixed at zero as our final values, but quote the parameters from both cases in Table  \ref{finalresults:orbit}, which contains the values of the orbital parameters for the case of non-zero eccentricity (eccentric) and eccentricity fixed at 0 (circular).

\begin{figure}
\plotone{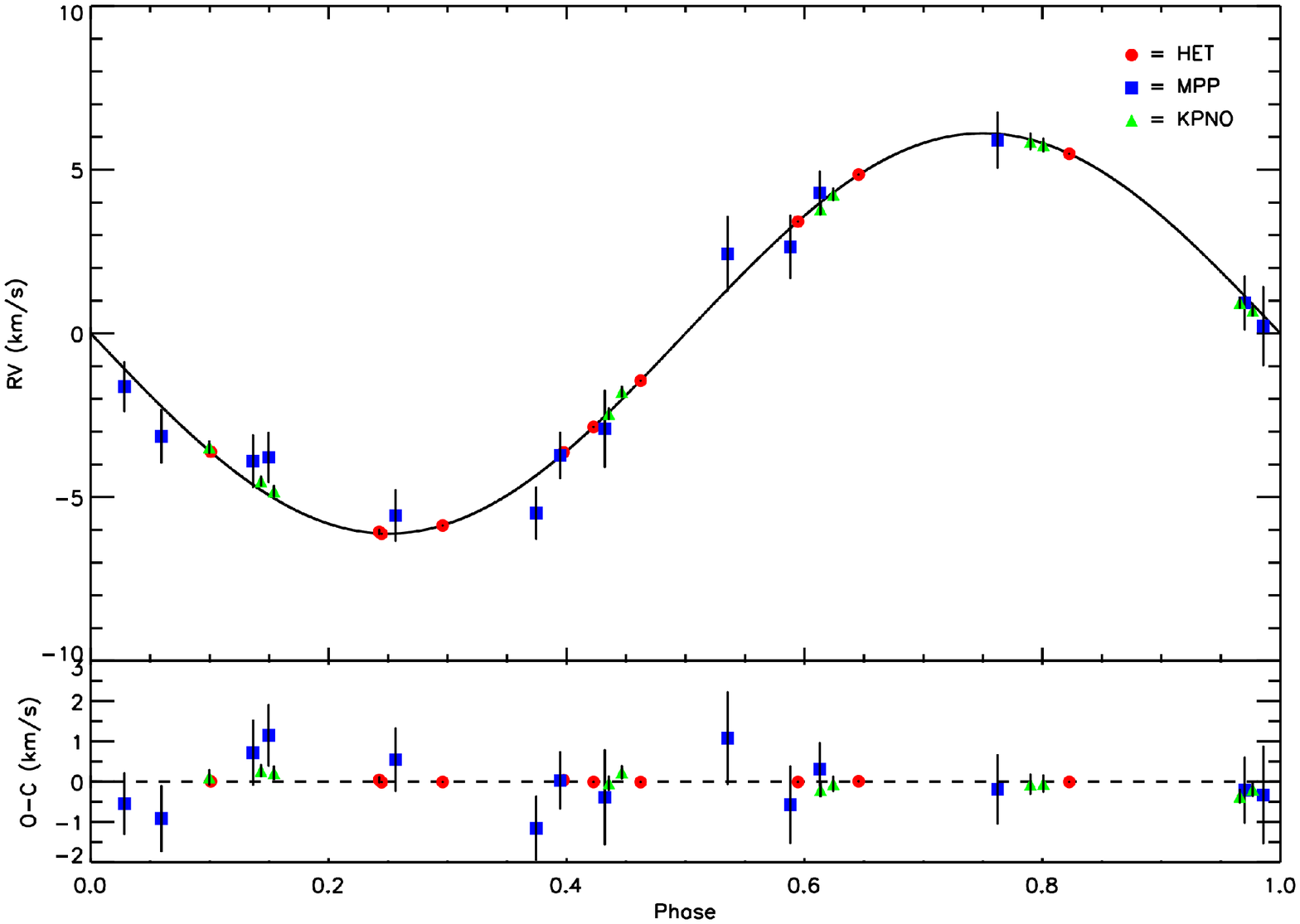}
\caption{Results from the combined MPP, KPNO and HET analysis, where the MPP errors are scaled by a factor of 8.49, KPNO errors are scaled by a factor of 3.37 and the HET errors are scaled by a factor of 0.3.  The eccentricity is fixed at $e = 0$.  MPP data are in blue, KPNO data are in green and HET data are in red.  The systemic velocity $\gamma_{0}$ has been removed.\label{finalfit}}
\end{figure}

Fig.\ \ref{finalfit} shows the final results of the joint RV fitting and fixing the eccentricity at 0.  MPP data are the blue squares, KPNO data are the green triangles, HET data are the red circles, and the systemic velocity $\gamma_{0}$ has been removed.  We find $\gamma_{0} = 18.68 \pm 0.24 ~ \rm{km~s^{-1}}$ for the star, with an offset between the MPP and KPNO data of $14.90 \pm 0.25 ~ \rm{km~s^{-1}}$ and an offset between the MPP and HET data of $12.61 \pm 0.24 ~ \rm{km~s^{-1}}$. The final orbital period is determined to be $5.69449 \pm 0.00023$ days and an RV semi-amplitude of 6.113 $\pm$ 0.009 km s$^{-1}$.  We searched for an additional signal in the residuals from the joint fit that might be caused by an additional companion in the system, but found no other frequencies with significant power.

\section{Relative Time Series Photometry of the Host Star}

Photometric observations are an important step in analyzing low-mass
companions discovered via the Doppler technique.  High-precision
photometry can be used to search for transits of the companion.
Additionally, time-series photometry can be used to rule out stellar
mechanisms of Doppler variability, such as chromospheric activity due
to starspots or stellar pulsations.  In the case of stars with
detectable starspots, time-series photometry can be used to determine
a stellar rotation rate.  In this section we present and analyze 
time series relative photometry of TYC-2949-00557-1
from two sources: relatively precise (few mmags) photometry
covering a relatively short timespan (2-8 hours over five nights)
from the Hereford Arizona Observatory,
and less precise (few percent), but more comprehensive photometry
consisting of 7194 epochs taken over roughly three years as part of the
Kilodegree Extremely Little Telescope (KELT) North transit survey.
Neither datasets show any evidence for variability of the host star.

\subsection{Relative Photometry from Hereford Arizona Observatory}

Initial photometric observations of the primary were performed on four
nights in 2009 (2/19, 2/21, 2/27, and 3/16) at the Hereford Arizona
Observatory (observatory code G95 in the IAU Minor Planet Center), a
private facility in southern Arizona.  Additional observations were
made in 2010 on 4/15 to search for transits based on an updated transit ephemeris from the
combined RV analysis (\S\ref{combined}).  All data were taken with an 11-inch
Celestron Schmidt-Cassegrain (model CPC 1100) telescope that is
fork-mounted on an equatorial wedge, an SBIG ST-8XE CCD with
a KAF 1602E detector, and an SBIG AO-7 tip-tilt image stabilizer used to
maintain the field at a fixed position on the CCD.  The observations
in 2009 were done without a filter (``C'' band), resulting in an effective
central wavelength of $\sim$570 nm between Johnson $V$ and $R$
bands.  The observations in 2010 were taken with a Sloan $r^\prime$
filter. Data toward the end of the night on 4/15/2010 were taken at very high
airmass (out to $\sec{z} = 5.7$), resulting is somewhat degraded photometric
precision. 

\begin{figure}
\plotone{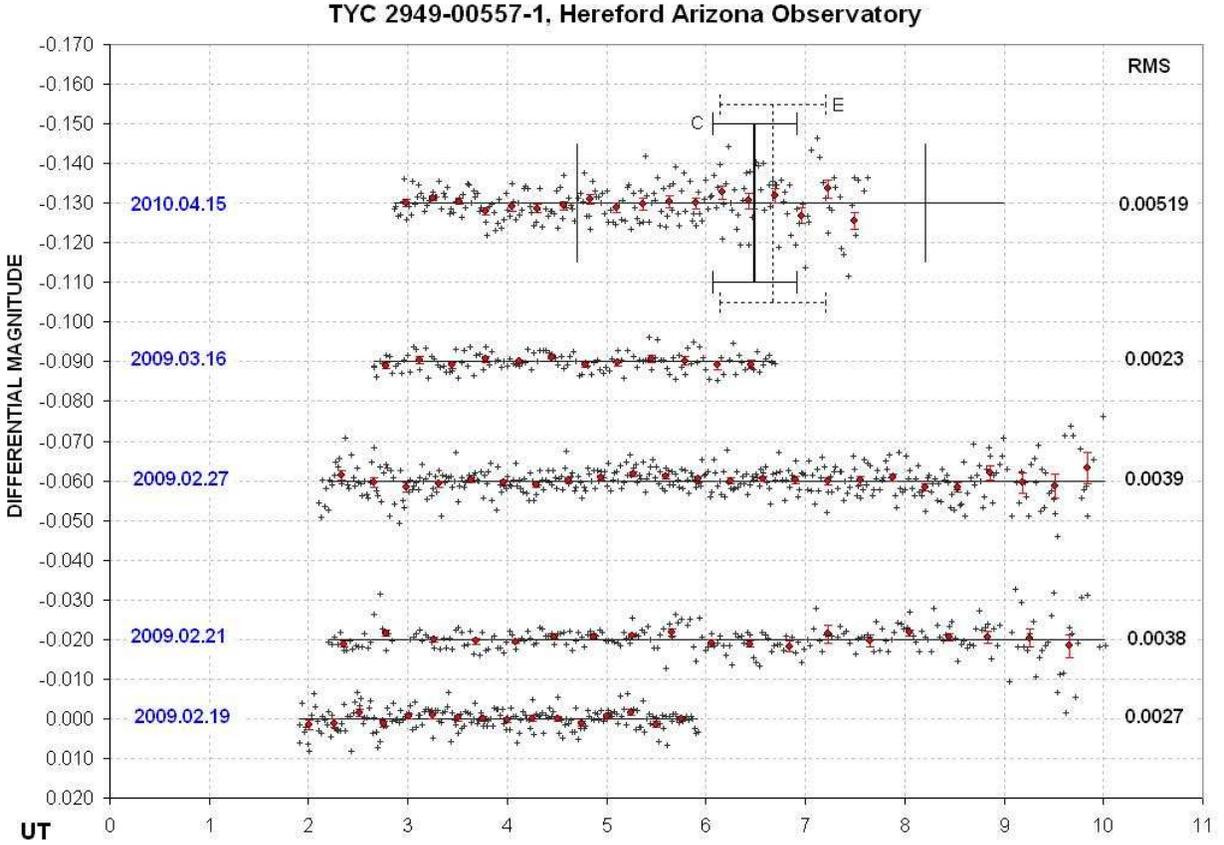}
\caption{Photometric observations with the Hereford Arizona Observatory telescope.  The horizontal axis is elapsed time each night in hours.  The star is photometrically stable with an RMS of 2-4 mmag.  The increased scatter towards the end of 2009.02.21, 2009.02.27 and most of 2010.04.15 are due to observing at high airmass.  The vertical bars are the predicted times of ingress, mid-transit and egress based on the RV fit in \S\ref{combined} and an assumed transit duration of 3.3 hours.  The two mid-transit estimates are based on the two results from the RV fitting (``C'' is for $e = 0$, ``E'' is for the case where $e$ is left as a free parameter).  The widths correspond to the uncertainties in the mid-transit times.\label{garyobs}}
\end{figure}

Fig.\ \ref{garyobs} shows the relative photometry over the five
nights, which demonstrate that the primary star is intrinsically
stable on the time scale of several hours, at the level of 2-4 millimagnitudes.  
Based on the final ephemerides determined in \S\ref{combined}, only
the last night (4/15/10, top row) covers possible times of predicted
transits.  The vertical bars are the predicted times of ingress,
mid-transit and egress based on the RV fit in \S\ref{combined}  for the
two methods of RV fitting (``C'' is for $e =
0$, ``E'' is for non-zero eccentricity), and an
assumed transit duration of 3.3 hours, corresponding to a nearly central
transit.  The widths correspond to the
uncertainties in the mid-transit times.  There is no
evidence of a transit at the most likely depth ($\sim 0.8\%$) and
duration during these observations.  
In \S\ref{sec:translimits} we consider the
uncertainties in the ephemeris and properties of the primary,
as well as a range of impact parameters, 
to quantify the confidence with which we can
exclude transits in this system.

\subsection{Relative Photometry with KELT}

The Kilodegree Extremely Little Telescope (KELT) North survey is a
wide-field photometric survey of $\sim 40\%$ of the northern sky
designed to monitor fairly bright ($8<V<12$) stars in order to search
for planetary transits \citep{KELT_THEORY,KELT_SYNOPTIC,siverd09}.
The KELT survey instrument consists of an Apogee AP16E (4K x 4K
9$\mu$m pixels) CCD camera attached to a Mamiya 645 medium-format 42mm
aperture camera lens.  The resultant field of view is $26^\circ
\times 26^\circ$ at roughly 23$^{\prime\prime}$ / pixel.  
The standard configuration uses a Kodak Wratten \#8
red-pass filter and the resultant bandpass resembles a widened
Johnson-Cousins $R$-band.  KELT-N is permanently mounted on
a fixed pier at Winer Observatory in Sonoita, AZ.

The KELT-N survey targets 13 star fields centered at $31.7^\circ$
declination (the survey site latitude) spaced fairly evenly through
all 24 hours of R.A. with slight overlap.  Exposure times are $150$
seconds, which yields relative photometric precisions of better than a
few percent for $V\la 12$. Typical cadences are roughly 20 to 30
minutes when the target field is visible, and to date there exist
$\sim 5000-7000$ epochs per target.  The areal sky coverage, target
magnitude range, and high photometric precision of the KELT
survey results in excellent synergy with the MPP (as well as the full
MARVELS survey).  TYC 2949-00557-1 is in one of KELT's target fields and is a good example of this synergy.  We use the KELT photometry to characterize the photometric variability of the host star, and to search for signatures of transits
of the companion.  We first briefly describe the data reduction, and
then describe the light curve analysis. 

Images of the field are flat-fielded,
and then relative photometry is extracted using the ISIS image subtraction
package \citep{ISIS}, in combination with DAOPHOT \citep{daophot} to
perform point-spread function fitting photometry.  We further
eliminate problematic images due to poor observing conditions by
examining outliers from the ensemble of individual light curves on the
CCD.  Any epochs that produce photometric outliers in a significant
fraction ($\ga 5\%$) of light curves are removed from all light
curves.  We use the VARTOOLS program \citep{hartman08} to remove
common trends due to systematic errors from the light curves using
the Trend Filtering Algorithm \citep{tfa}, first removing the 20 points with the highest flux and the 20 points with the lowest flux from all
light curves.  We choose the 400 stars with the lowest RMS values as
comparison stars for trend removal, ensuring that these stars are
evenly distributed across the region of the CCD near the target, and
excluding variable stars and saturated stars.  Finally, the errors of
all the light curves are scaled by a constant value, chosen to force the modal value of 
$\chi^2/{\rm dof}$ for a weighted, constant flux fit to the light curves
to be unity for the ensemble.
For TYC 2949-00557-1, this procedure resulted in a $\chi^2/{\rm dof}$ that
differed somewhat ($\la 50\%$) from unity.  Since
we see no evidence for photometric variability for this star (see below)
we further scale the errors to force $\chi^2/{\rm dof}=1$ to be conservative.
TYC 2949-00557-1 happened to fall in an overlap region of two target
fields, and as a result we had two sets of light curves for the target. These
data were reduced independently, and then combined after subtracting the difference
between their weighted mean magnitudes. 

\begin{figure}
\plotone{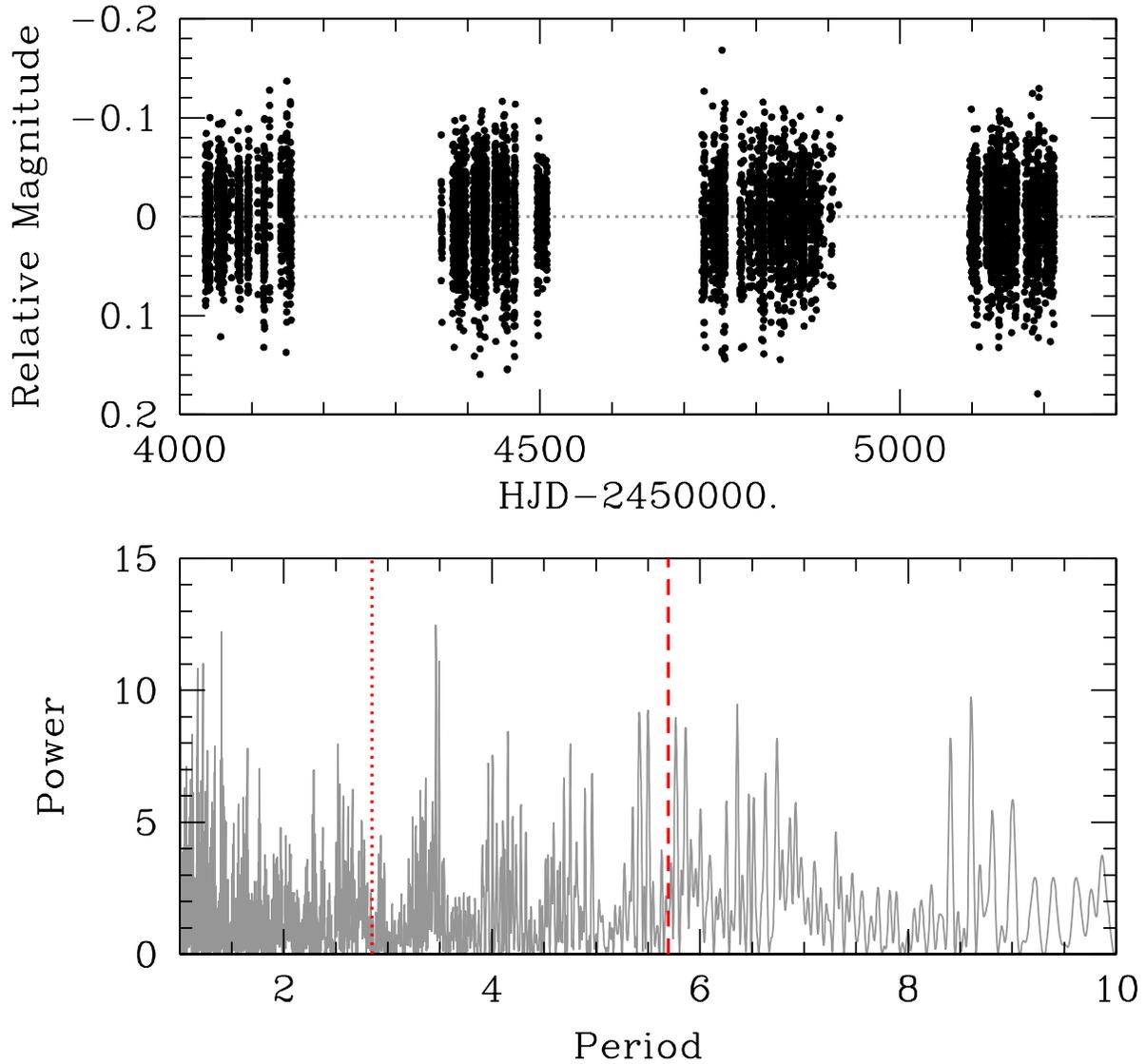}
\caption{
(Top panel) KELT-N light curve for TYC 2949-00557-1.  (Bottom panel)
Lomb-Scargle periodogram of the KELT data,
showing no evidence for any significant periodicities
for periods of $P=1-10~{\rm days}$, including the period of the RV companion (vertical dashed line) and the first harmonic (vertical dotted line).
\label{kelt}}
\end{figure}

\begin{figure}
\plotone{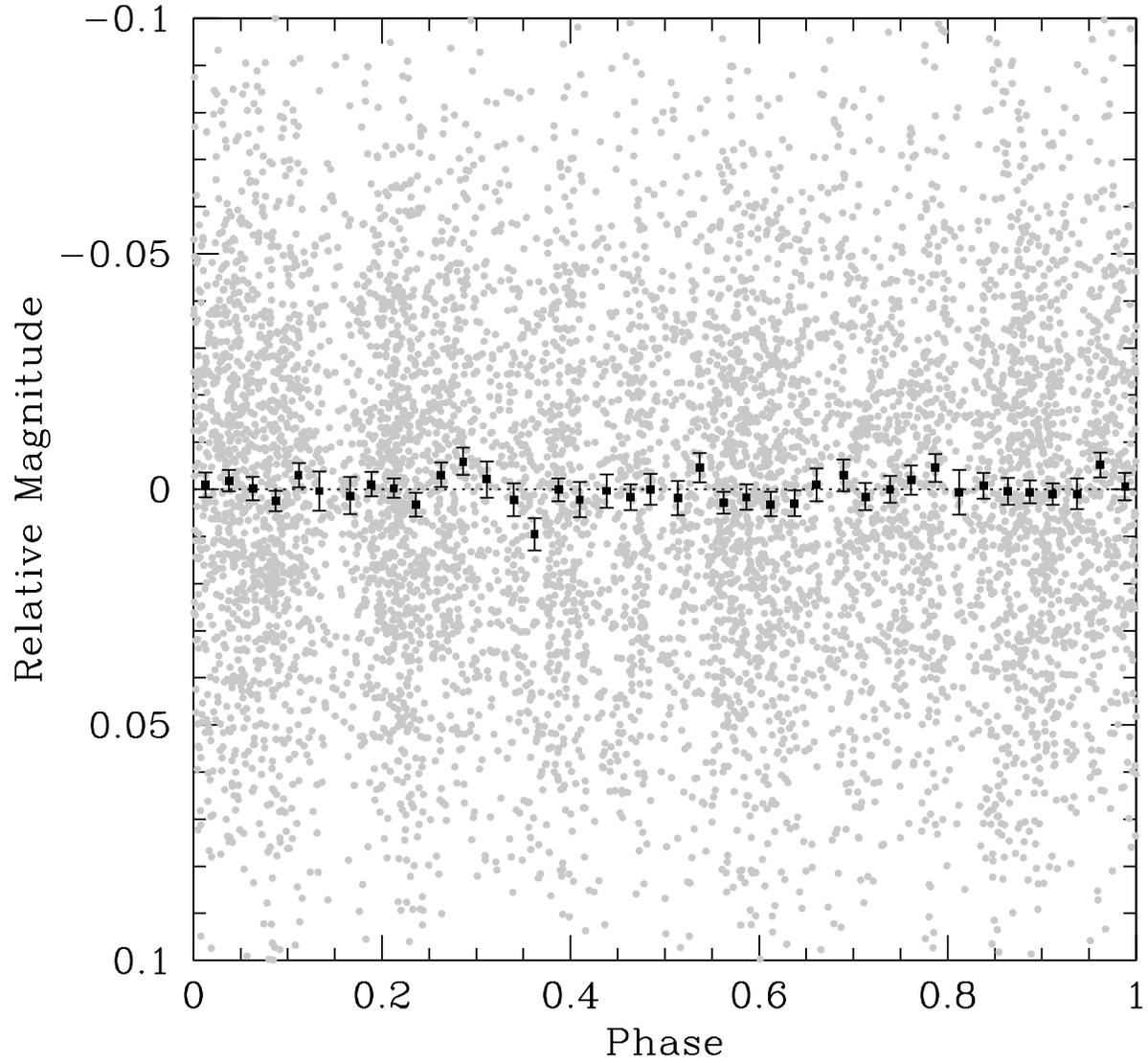}
\caption{
The small filled circles show the KELT-N light curve for TYC 2949-00557-1 
phased according to the ephemeris from the joint
RV fit (see \S\ref{combined}). The larger filled squares with error
bars are binned 0.025 in phase. 
\label{binkelt}}
\end{figure}

The KELT lightcurve for TYC 2949-00557-1 is shown in Fig.\ \ref{kelt}.
It contains 7194 data points spanning 3.2 years, and has a weighted
RMS of 3.5\%.  We search for variability using a weighted Lomb-Scargle
periodogram with floating mean \citep{lomb76,scargle82,cumming99}, and
find no significant peaks (see Fig.\ \ref{kelt}), and in particular
no evidence for periodic variability near the period
of the companion $(P \simeq 5.69$ days), or the first harmonic ($P/2$).
Fig.\ \ref{binkelt} shows the KELT lightcurve, phased according
to the best-fit RV ephemeris (Table \ref{finalresults:orbit}), as
well as binned $0.025$ in phase (3.46~{\rm hrs}).  The RMS of the binned data is $\sim 2.4$ mmag, and the $\chi^2/{\rm dof}=0.91$ for
a constant fit, indicating a low level of correlated noise, and
no evidence for variability at the few mmag level.  
We limit the amplitude of any variability at $P/2$
to be $\la 2$ mmag; unfortunately this
is well above the level of ellipsoidal variability expected for
this companion of $\sim 0.03 ~ \sin i~{\rm mmag}$ \citep{pfahl08}.

\subsection{Limits on Transits\label{sec:translimits}}

Given the relatively high \emph{a priori} transit probability of TYC 2949-00557-1b of
$\sim R/a \sim 8\%$, where $R$ is the stellar radius, we searched for transits in the KELT dataset
combined with the Hereford data from 4/15/10.
The expected transit duration is  $\sim RP/(\pi a) \sim 3.4~{\rm hrs}$
for a central transit,
and the expected fractional depth is $\delta \sim (r/R)^2 \sim 0.8\% ~ (r/R_J)^2$, where $r$ is the radius of the companion.  The
expected radius
of the companion depends on its true mass, as well as the
age of the system, but is likely to be $\sim 1 ~ R_J$ \citep{baraffe03}. 
Unfortunately, while the KELT data has excellent phase coverage, it is not of sufficient quality to 
detect or rule out the expected signal.  
If transits were present, we should expect to detect them
with a signal-to-noise ratio of 
\begin{equation}
{S/N} \sim N^{1/2} \left(\frac{R}{\pi a}\right)^{1/2} \frac{\delta}{\sigma}
\label{eqn:transitsnr}
\end{equation}
where $N=7194$ is the number of data points, and
$\sigma \sim 3.5\%$ is the typical uncertainty. Thus, $S/N \sim 3 (r/R_J)^2$, which is marginal unless the companion has a radius significantly larger than Jupiter.  On the other hand, the majority of the Hereford data is generally of sufficient
quality to detect transits at the expected depth, and one night covers 
the predicted transit time for the companion.  Unfortunately,
there is no indication of a transit
at the expected time.

We nevertheless proceed with a quantitative search for a transit signal.
We combine the KELT data with the Hereford data from 4/15/10 after first subtracting the difference between
their weighted mean magnitudes.  The slight difference in passband between the HAO and KELT data do not affect the ability to detect a transit in the combined data set.  We use the distribution of companion periods $P$ and expected transit
times $T_c$ from the MCMC analysis of the combined radial velocity
data described in \S\ref{combined}.  For each combination
of $T_c$ and $P$ (i.e., for each link in the Markov Chain), we 
draw a random value for the $T_{\rm eff}$, $\log{g}$, and [Fe/H] of the 
primary from a Gaussian distribution, with the central values and
dispersions given in Table \ref{finalresults:mags}, as determined from the spectroscopic
analysis described in \S\ref{specfollow}.  We then use the
\citet{tor2010} empirical relations to estimate the mass $M$
and radius $R$ of the primary for those values.  We add an additional
offset to $M$ and $R$ drawn from Gaussians with dispersions equal
to the dispersions of the fits to the empirical relations in
\citet{tor2010}; specifically $6.4\%$ in $M$ and $3.2\%$ in R. Finally,
we draw a random value of the impact parameter of the transit in units
of the radius of the star in the range [0,1].  Assuming a radius
for the companion, we then compute the
expected transit curve using the routines of \citet{mandel02},
using limb-darkening coefficients from \citet{claret00},
assuming that both the KELT and Hereford bandpasses roughly correspond to $R$.  We
then fit this curve to the combined dataset, and compute the improvement
in $\chi^2$ relative to a constant flux fit to the data.  We repeat
this for each link in the Markov chain, as well as for a variety
of different companion radii. 

We search for significant improvements in $\chi^2$ which would
be indicative of a detection.  Our best-fit has $\Delta\chi^2=-11.7$
relative to a constant flux fit. In order to asses the
significance of this improvement in $\chi^2$, we repeat the search for ``anti-transit'', i.e., signals with the same shape as a transit but
corresponding to an increase in flux (see \citealt{burke06}), and find
improvements in $\chi^2$ at similar levels.  We therefore conclude that
there is no evidence for a transit in the combined KELT and Hereford data.

\begin{figure}

\plotone{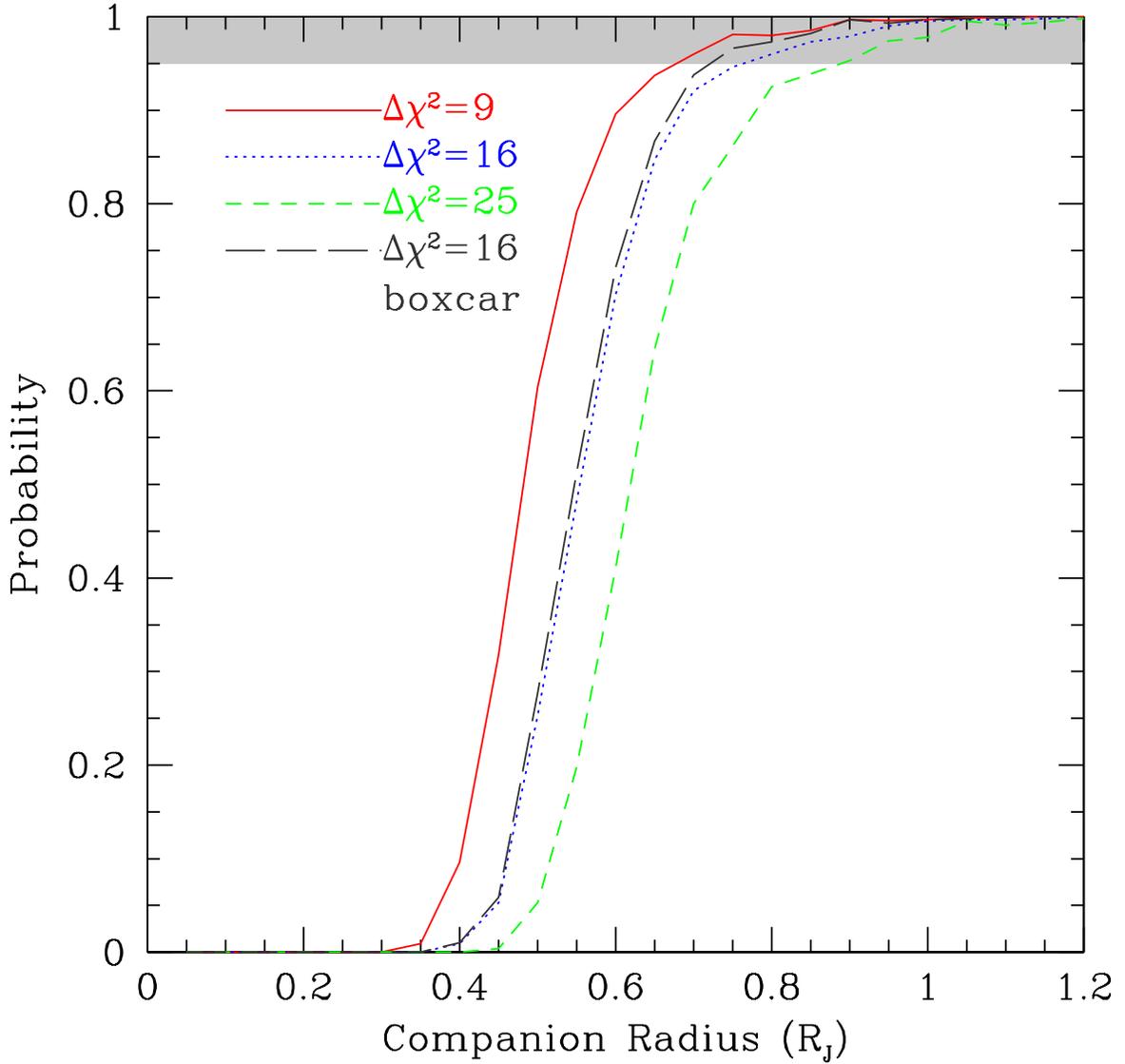}
\caption{
Probability that transits of a companion are excluded
at levels of $\Delta\chi^2=\{9,16,25\}$ based on the analysis of the combined Hereford and KELT photometric datasets, as a function of the radius
of the companion.  The black, long-dashed line is a case where boxcar-shaped transits were used as a test, and is for $\Delta\chi^2=16$. Transits of companions with radius $r\ga 0.75~R_J$ can be excluded
at the $95\%$ confidence level.
\label{exctrans}}
\end{figure}

Given that we have not detected any evidence of transits, we now ask
what the probability is that we would have detected a transit of a
given radius, {\it assuming} that the companion transits (i.e., $b\le 1$).  
To do this, we simply determine what fraction of the steps in the
Markov Chain described above result in an increase in
$\chi^2$ above a certain level, as a function of the radius of the
companion.  This result is shown in Fig.\ \ref{exctrans}, for
$\Delta\chi^2=\{9,16,25\}$.  The $\Delta\chi^2$ values were chosen as representative values: $\Delta\chi^2 = 16$ is the likely detection limit, 25 is chosen as a conservative limit and 9 is chosen to straddle the true detection limit.  The black, long-dashed line is a case where flat-bottomed, boxcar-shaped transits (no ingress/egress) were used and represents $\Delta\chi^2=16$.  It shows that detailed modeling of limb darkening and the ingress/egress has little effect on the final results of this test.  Given the properties of the noise as
estimated from the improvements in the fits from ``antitransits'',
signals with $\Delta\chi^2 \ga 16$ are likely to have been
reliably detected.  Thus, we can exclude transits of a companion with $r \ga 0.75 ~ R_J$.  
at the 95\% confidence level.
\citet{baraffe03} predict radii of $\ga 0.77 ~ R_J$ for brown
dwarfs with $m \simeq 60~M_J$ for ages of $\la 5$~Gyr.  We conclude
it is unlikely this companion transits, unless the system
is substantially older than 5 Gyr, which is unlikely given
the effective temperature and surface gravity of the host star (\S\ref{hostprop}).

\section{Absolute Photometry}

The Tycho-2 catalog's \citep{hog2000} $V$-band magnitude for this object
is 11.840, however, Tycho magnitudes are known to significantly degrade beyond
$V_{T} > 11.$ Measurements were taken at Hereford Arizona Observatory using
both $B,V,R_c,I_c$ and Sloan $g^\prime,r^\prime,i^\prime$ filter
sets.  A total of 64 Landolt standard stars and 15 SDSS standard stars were used as calibrators.  The
conversion equations of \citet{smi2002} were used to convert the Sloan
filter measurements into the $BVRI$ system.  The agreement between the
observed $BVRI$ magnitudes and those converted from Sloan magnitudes
agree within the measurement error.  We adopt the unweighted average
of each $BVRI$ measurement and the larger of the two statistical errors
for the final magnitude results.  Table \ref{multimagtable} summarizes
the measured magnitudes and final results for the multi-band
photometry.  Near-IR fluxes are taken from the 2MASS \citep{skr2006}
Point Source Catalog and are presented in Table \ref{finalresults:mags}.

\section{Characterization of the Host Star}
\subsection{SED Analysis \label{sedfitting}}

\begin{figure}
\plotone{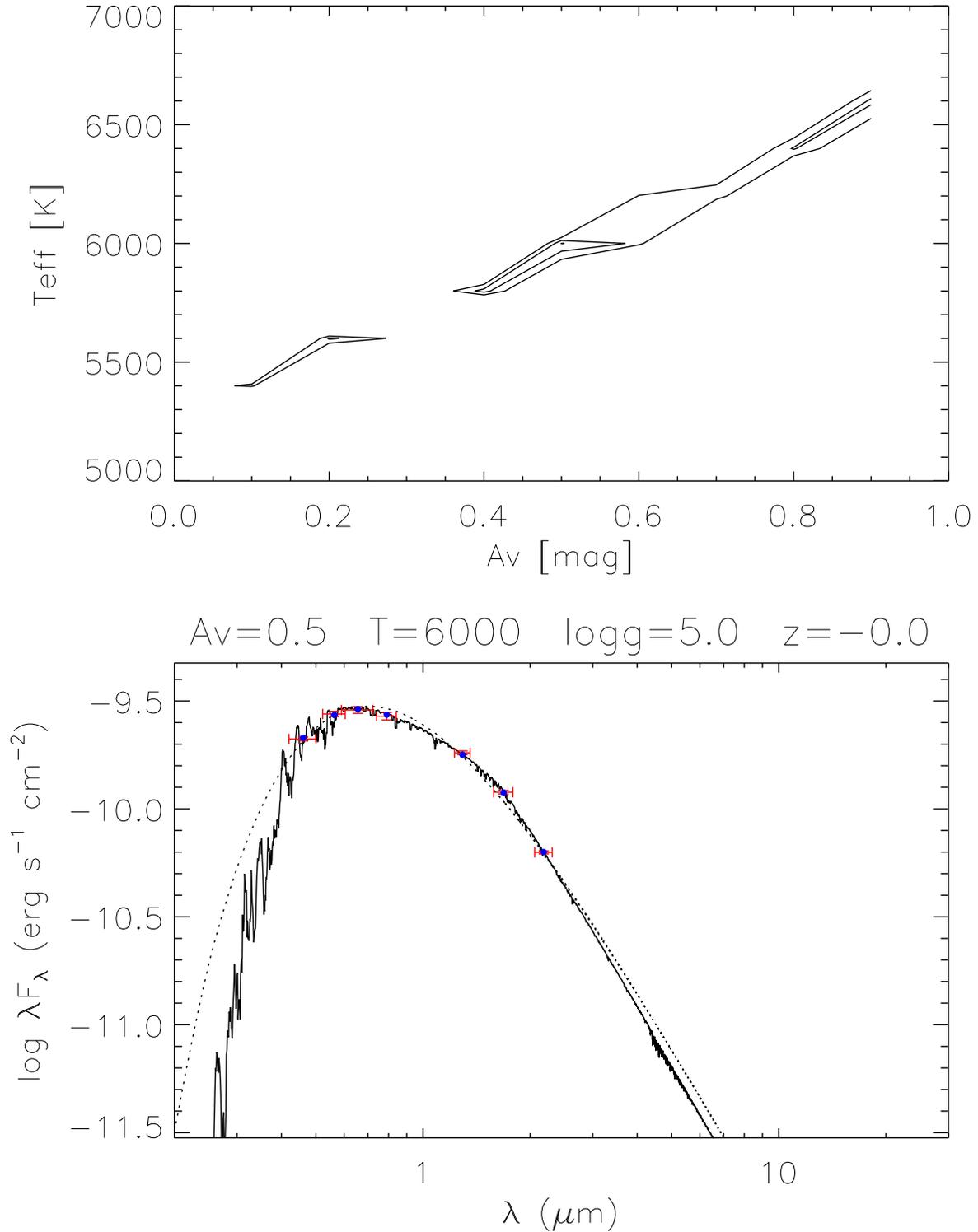}
\caption{Top: ${\chi}^2$ map in $T_{\rm eff}-A_{V}$ space showing the degeneracy between extinction and $T_{\rm eff}$. Inner contours are 1-$\sigma$ uncertainties for $T_{\rm eff}$ and $A_{V}$ assuming they are the only significant degrees of freedom, outer contours assume all four parameters are significant.  Bottom:  NextGen model overplotted on the observed fluxes.  The dotted curve is a blackbody SED for the best-fit $T_{\rm eff}$.\label{nextgensed}}
\end{figure}

We use the $BVRI$ fluxes along with the 2MASS near-IR data to fit model spectral energy diagrams (SEDs) to derive approximate stellar parameters.  NextGen models from \citet{hau1999} in grids of 100 K for $T_{\rm eff}$, 0.5 dex for
$\log g$ and 0.5 dex for the metallicity, represented by [Z/H], are fit
along with the line-of-sight extinction.  There is a degeneracy
between line-of-sight extinction and derived $T_{\rm eff}$ when fitting
SEDs with unknown extinction.  Fig.\ \ref{nextgensed} (top) shows the
$\chi^2$ map in $T_{\rm eff}-A_{V}$ space.  The interior contours
represent 1-$\sigma$ uncertainties assuming that $T_{\rm eff}$ and $A_{V}$
are the only two free parameters.  The exterior contours are the
1-$\sigma$ uncertainties with all four parameters ($T_{\rm eff}$, $A_{V}$,
$\log g$ and [Z/H]) as free-parameters.  We find the global minimum in
$\chi^2$-space is for a $T_{\rm eff} = 6000$ K, $\log g$ of 5.0, [Z/H] of
0.0 and extinction $A_{V}$ = 0.5.  Fig.\ \ref{nextgensed} (bottom)
shows the best-fit NextGen model along with the observed fluxes.  Other
solutions exist within the 1-$\sigma$ contours at cooler $T_{\rm eff}$ and
smaller $A_{V}$; however, the dust maps of \citet{sch1998} give an
$A_{V}$ of 0.45 along this line-of-sight, consistent with the best-fit
$A_{V}$ of 0.5 assuming that the star is behind the majority 
of the dust along this line of sight. The hotter temperature solution is also consistent
with the $T_{\rm eff}$ derived from Echelle spectra presented in \S\ref{specfollow}.

\subsection{Spectral Synthesis\label{specfollow}}
In order to derive physical properties of the host star and estimate
the minimum mass of the companion, stellar templates from the HET
observations that do not contain iodine lines were used to derive
parameters of the host star.  We use the latest MARCS model
atmospheres \citep{gus2008} for the analysis. Generation of synthetic
spectra and the line analysis were performed using the turbospectrum
code \citep{alv1998}, which employs line broadening according to the
prescription of \citet{bar1998}. The line lists used are drawn from a
variety of sources. Updated atomic lines are taken mainly from the Vienna Atomic Line
Database (VALD) database \citep{kup1999}. The molecular species CH, CN, OH, CaH,
and TiO are provided by B. Plez \citep[see][]{ple2005}, while the NH,
MgH, and C2 molecules are from the Kurucz linelists. The solar
abundances used here are the same as \citet{asp2005}.  The FeI and
FeII lines used for the line analysis were compiled
by \citet{san2004}, who use solar Fe abundance to derive gf
values.  Both \citet{asp2005} and \citet{san2004} use HARPS solar spectra and an Fe abundance of 7.45.

\begin{figure}
\plotone{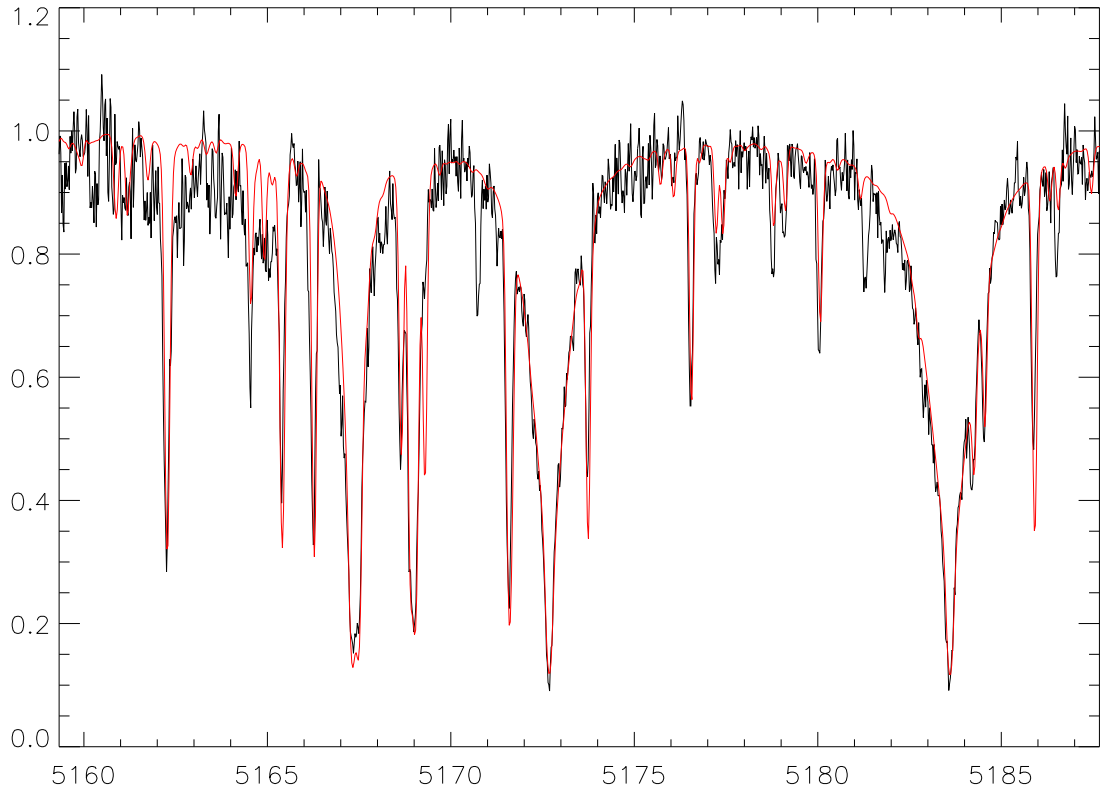}
\caption{HET/HRS template spectrum in black with best-fit model in red for the Mgb region.  The spectra have been continuum normalized and the relative flux density is plotted against the wavelength in Angstroms.\label{hetmgb}}
\end{figure}

We derive a $T_{\rm eff}$ = 6135 $\pm$ 40 K and [Fe/H] = 0.32 $\pm$ 0.01,
based on FeI excitation equilibrium and a $\log g$ = 4.4 $\pm$ 0.1
based on the ionization equilibrium of FeI and FeII lines and by
fitting the wings of the Mgb triplet at 5167, 5172 and 5183 \AA.  The
error estimates are based on the equivalent width of Fe lines and the
errors of Fe abundances from the individual lines.  A microturbulence
value ${\xi}_{t}$ = 1.65 km s$^{-1}$ is derived by forcing weak and strong
FeI lines to give the same abundances. Fitting the Fe lines in the Mgb
region yields a rotational velocity of $v\sin I$ = 7 km s$^{-1}$. We only
used the 110 FeI lines weaker than 100 m\AA\ for the analysis.
Fig.\ \ref{hetmgb} shows the continuum normalized spectra in black and
the best-fit model in red for the Mgb region.

In addition to the HET templates, we obtained high-resolution (R
$\sim$ 31,000) Echelle spectra using the ARCES
instrument \citep{wang03} on the APO 3.5m telescope.  Seven exposures
for a total of 63 minutes of integration were obtained.  Data was
reduced using a modified IRAF script originally written by
J. Barentine and J. Krzesinski for ARCES data.  Spectra are corrected
for bias and dark subtraction, cosmic rays and bad pixels.
Flatfielding is performed using a quartz lamp and two different sets
of integration times: a ``blue'' set of 4-min integrations using a
blue filter and a ``red'' set of 7-sec integrations with no filter in
the beam.  These sets are then combined to form a master flatfield
image.  The two different quartz sets are used to maximize the
signal-to-noise in both the blue and red end of the spectrum.  Spectra
are wavelength calibrated using a sequence of 10-sec ThAr integrations
taken a few times during each night.  The star HD 42088 (spectral type
06V) was observed to remove telluric lines.

The spectrum was analyzed using the IDL-based program Spectroscopy
Made Easy, or SME \citep{val1996}. This code uses synthetic spectra
and multidimensional least-squares minimization to determine the best set of stellar parameters for an
observed spectrum. These parameters
include effective temperature, surface gravity, metallicity,
microturbulence, macroturbulence, projected rotational velocity, and
the radial velocity. We follow the guidelines from previous
spectroscopic studies of host stars that used
SME \citep[e.g.,][]{val2005, ste2007}.  We used three-dimensional
interpolation on the \citet{kur1993} grid of local thermodynamic
equilibrium (LTE) model atmospheres and the VALD database to obtain line data for transitions with predicted
absorption cores deeper than $0.5\%$ of the continuum.  For the VALD
queries, we used solar abundances, $T_{\rm eff} = 5770$ K, $\log g$ = 4.44
and ${\xi}_{t} = 0.866$ km s$^{-1}$.

\begin{figure}
\plotone{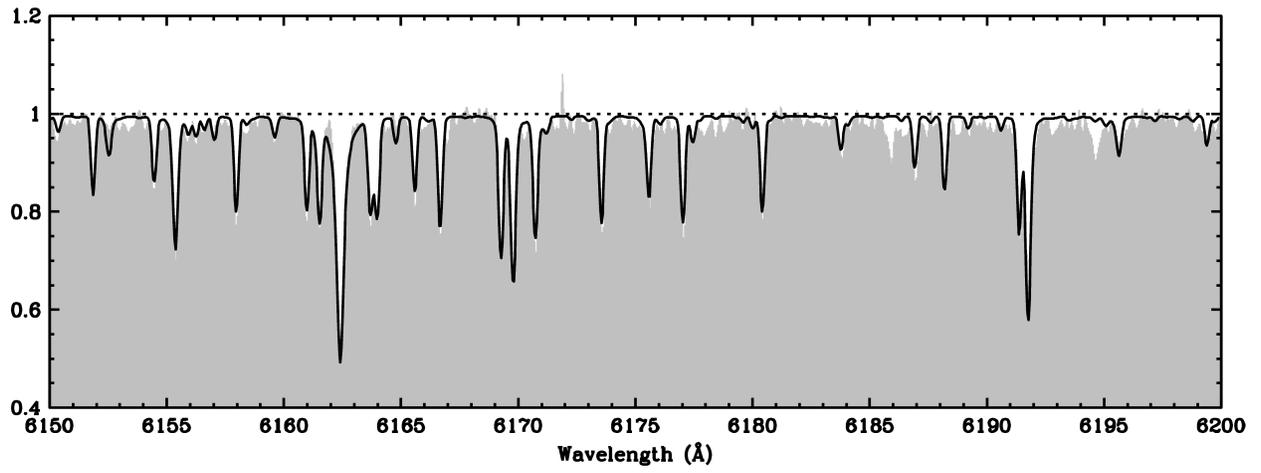}
\caption{A portion of the ARCES spectrum (R $\sim$ 31,000) used in the SME analysis.  The input spectrum is in white with the best-fit model overlaid in black.\label{smeplot}}
\end{figure}

For the SME analysis, as suggested by \citet{ste2007}, we fixed the parameter ${\xi}_{t}$ to 0.85 km s$^{-1}$,
in order to decouple
the correlation between microturbulence ${\xi}_{t}$ and metallicity.
For the macroturbulence
${\zeta}_{t}$, we follow the empirical relation of \citet{val2005}
which gives ${\zeta}_{t} = 4.5$ km s$^{-1}$ for a star with $T_{\rm eff} \sim
6200$ K.  We were unable to obtain consistent results from the
gravity-sensitive Mgb triplet region, therefore we fix $\log g$
at three values of 4.3, 4.4, and 4.5, corresponding to the 1-$\sigma$ range determined
from the HET spectra.  We set as free parameters $T_{\rm eff}$, [M/H],
$v \sin I$, and the radial velocity $v_{rad}$ and utilize the
metal-rich region of $6000 - 6200$~\AA.  The uncertainties of the
parameters are derived from the range of best-fit results using the
three fixed $\log g$ values.  We derive $T_{\rm eff} = 6246 ^{+27} _{-45}$
K, [M/H] = $0.3615 ^{+0.009} _{-0.027}$ and $v\sin I = 7 \pm 1$
km s$^{-1}$, in reasonable agreement with the values derived from the HET
spectra. Fig.\ \ref{smeplot} shows the best-fit model in black and the
input spectrum in white for a portion of the $6000 - 6200$ \AA\ range
used in the fitting.

The SME analysis of the ARCES spectrum is used as an independent check
on the derived temperature and metallicity from the HET spectra.
Because it is based on a smaller wavelength region, and cannot be used
to independently derive the surface gravity, we chose to adopt the
results from the HET analysis for the final stellar properties.

\section{Determination of Host Star Mass and Radius\label{hostprop}}
An alternative to interpolating isochrone models to determine stellar properties is to apply the
analytical equations derived by \citet{tor2010} using measurements of
eclipsing binary systems.  We use this empirical relation to derive
the mass and radius of the primary using the values of $T_{\rm eff}$,
$\log g$ and [Fe/H] obtained from the HET template spectra.  Applying
the equations for $\log{M}$ and $\log{R}$ yields a mass of $M = 1.25 \pm 0.09
M_{\odot}$ and a radius of $R = 1.15 \pm 0.15 R_{\odot}$.
Correlations of the best-fit coefficients from \citet{tor2010} are
included in the errors, but correlations of $T_{\rm eff}$, $\log g$ and
[Fe/H] are not considered.  The reported scatter in the
relation as found in \citet{tor2010} of $\sigma_{logm} = 0.027$ and
$\sigma_{logr} = 0.014$ are also included in the mass and radius uncertainties, respectively, by adding them in quadrature.

\begin{figure}
\plotone{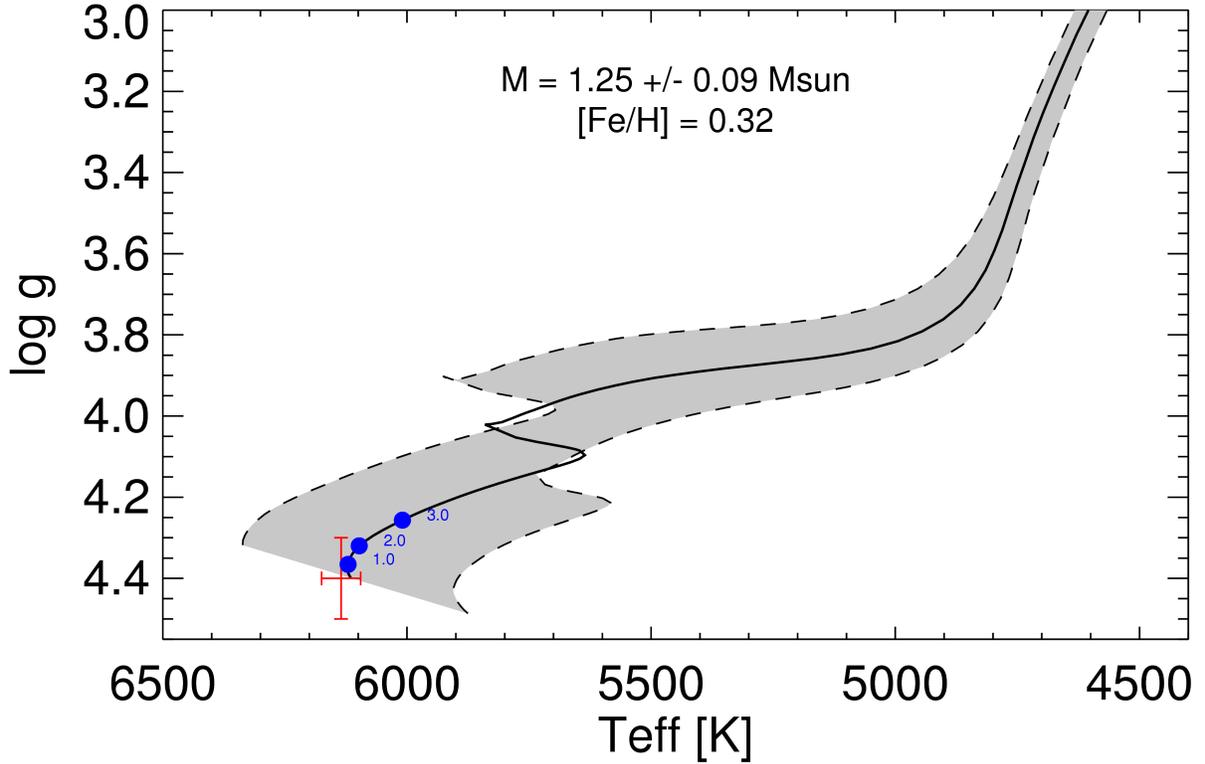}
\caption{HR diagram as a function of $T_{\rm eff}$ and $\log g$ based on Yonsei-Yale stellar evolution models.  The solid track is for the best-fit stellar parameters of 1.25 $M_{\odot}$ and [Fe/H] = +0.32.  The two dashed tracks are for masses of $1.25 \pm 0.09 M_{\odot}$ and represent the 1-$\sigma$ uncertainties on the mass.  Blue dots are the location of the star at ages of \{1,2,3\} Gyr, respectively.  TYC 2949-00557-1 is consistent with a ZAMS star that has an age no older than $\sim$ 2 Gyr.\label{tefflogghrd}}
\end{figure}

Fig.\ \ref{tefflogghrd} compares the spectroscopically measured
$T_{\rm eff}$ and $\log g$ of TYC 2949-00557-1 (red error bars)
to a theoretical stellar evolutionary track from the Yonsei-Yale
(``Y$^2$'') model grid \citep[see][and references therein]{dem2004}.
The solid curve represents the evolution of a single star of mass
$1.25\,{\rm M}_\odot$ and metallicity of [Fe/H]=$+0.32$, starting from
the zero-age main sequence (lower left corner), across the Hertzsprung
gap, and to the base of the red-giant branch.  Symbols indicate
various ages along the track labeled in Gyr.  The dashed curves
represent the same evolutionary track but for masses $\pm 0.09\,{\rm
M}_\odot$, representative of the $1\sigma$ uncertainty in the mass
from the \citet{tor2010} relation. The filled gray region between the
two mass tracks represents the range of expected locations of a star
like TYC 2949-00557-1 given the 1-$\sigma$ mass uncertainty and
measured metallicity. We emphasize that we have not directly measured
the mass of TYC 2949 but rather derived the mass using the empirical
relation of \citet{tor2010}.  Our purpose here is not to test the
accuracy of the theoretical stellar evolutionary tracks, but rather to
constrain the evolutionary
status of the TYC 2949 system.  The spectroscopically measured $T_{\rm
eff}$, $\log g$, and [Fe/H] place TYC 2949 near the zero-age main
sequence, with an age of at most $\sim$2 Gyr.

The distance to the host star can be computed once the bolometric luminosity is known.  We use the Stefan-Boltzmann law to derive the luminosity of the star using the $T_{\rm eff}$ found in \S\ref{specfollow} and the radius calculated by the \citet{tor2010} relation.  The absolute $V$ magnitude is then given by
\begin{equation}
M_{V} = -2.5 \log{\frac{L}{L_{\odot}}} + M_{Bol_{\odot}} - BC_{V}
\label{absv}
\end{equation}
where $M_{Bol_{\odot}}$ is the bolometric absolute magnitude of the Sun, here assumed to be 4.74, and $BC_{V}$ is the bolometric correction to the $V$ band.  We adopt a value of $BC_{V} = -0.175$, interpolated from Table 15.7 in \citet{cox2004} based on the $T_{\rm eff}$ of TYC 2949-00557-1.  The distance can then be calculated via the distance modulus and assuming a value for the line-of-sight extinction.  If we assume $A_{V} = 0.45 \pm 0.1$, which is
consistent with both the \citet{sch1998} dust maps and the best-fit value obtained from the SED fitting in \S\ref{sedfitting},  we derive a distance to the system of 413 $^{+66} _{-57}$ pc.  This distance is consistent with our implicit assumption that the star is behind the majority of the dust
along this line of sight, for likely values of the thickness of the dust layer. The quoted uncertainty in the distance includes the uncertainties in the stellar radius, effective temperature, line-of-sight extinction and apparent $V$ band magnitude.

\section{Companion Mass}

Given the orbital parameters from the joint radial velocity fit (\S\ref{combined}), 
and the estimate of the mass of the host star derived from the spectroscopic
parameters (\S\ref{specfollow}), we can estimate a minimum mass for the companion
of 
\begin{equation}
m_{\rm min}= 64.3 \pm 3.0 M_J,
\label{msini}
\end{equation}
where we have assumed a circular orbit.  This minimum mass is based only on the RV and stellar parameters,
and ignores the fact that edge-on configurations are likely excluded given the lack of transit
signature. Since the minimum mass is below
the hydrogen burning limit ($\sim 80~M_J$), one might be tempted to 
categorize this object as a brown dwarf.  However, since the orbital inclination $i$ is unknown, we do
not know the true mass, and given that the minimum mass is not far below the hydrogen
burning limit, it does not suffice to make the usual assumption that the minimum mass of
the object can be used to characterize its nature.  Rather, we must
be somewhat more careful to estimate the probability distribution for its true mass.

We proceed to estimate the probability distribution of the companion
mass using a similar method as was used to search for and exclude
transits in the KELT photometric data as described in \S\ref{kelt}. We
use the distribution of companion periods $P$, and velocity semi-amplitudes
$K$ from the Markov chain derived from the
MCMC analysis of the combined radial velocity data
described in \S\ref{combined}, for the fit assuming zero eccentricity. For each set of these parameters
(each link in the chain), we draw a random value for the
spectroscopically determined primary parameters $T_{\rm eff}$,
$\log{g}$, and [Fe/H], according to a Gaussian distribution centered
on the best-fit values and with dispersions equal to the uncertainties
(given in Table \ref{finalresults:mags}).  We use these values of $T_{\rm
eff}$, $\log{g}$, and [Fe/H] to estimate the primary mass $M$ using
the empirical relation of \citet{tor2010}.  We add an additional
offset to $M$ drawn from a Gaussian with dispersion equal to the
dispersion in the fit to this empirical relation (6.4\%).  We draw a
random value of $\cos{i}$ from a uniform distribution in the range
$(0,1)$, and then solve for the mass $m$ of the secondary (note we do
not assume that $m\ll M$).

\begin{figure}
\plotone{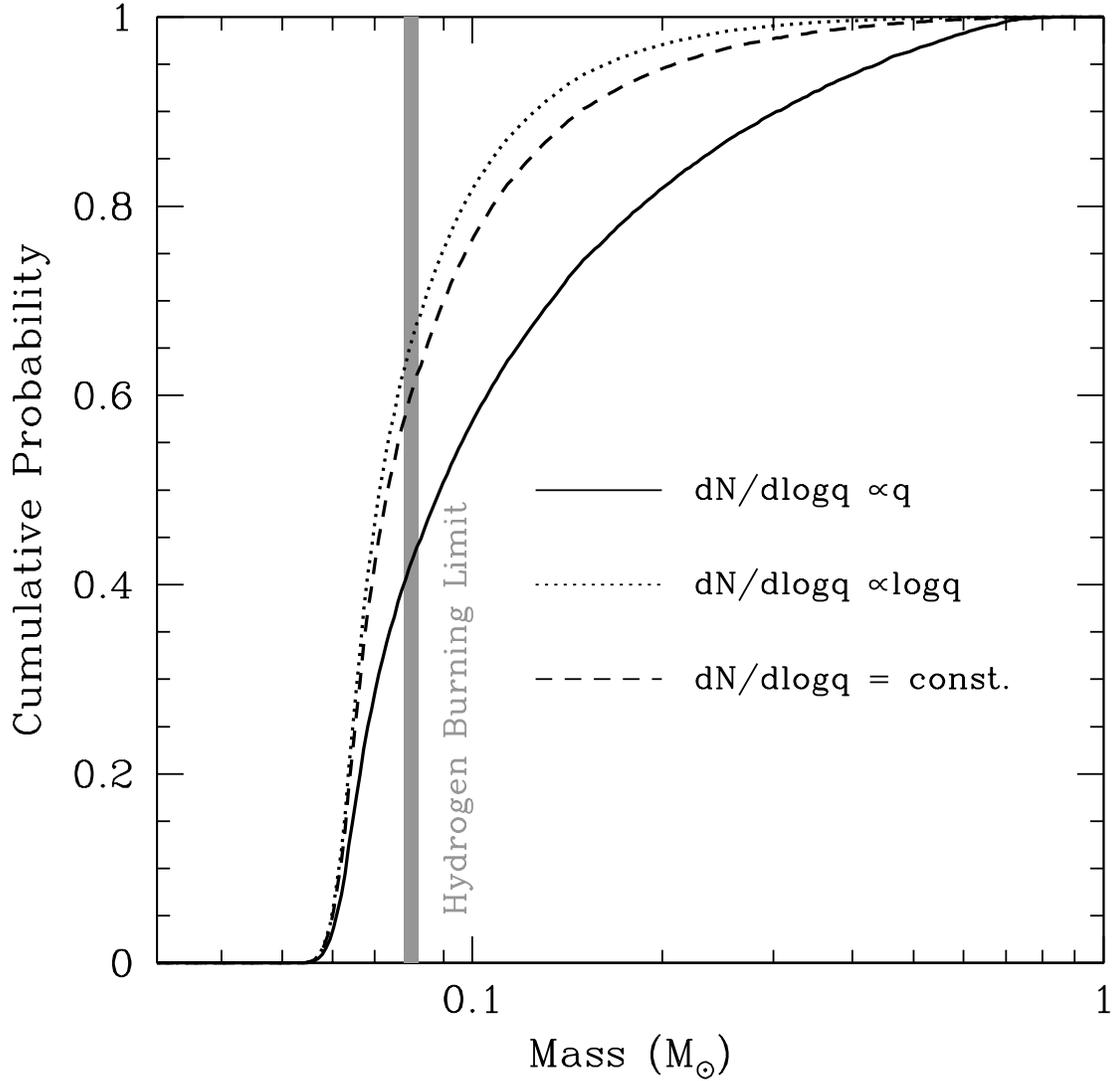}
\caption{Cumulative probability that the mass of the companion
to TYC 2949-00557-1 is less than a given mass in solar masses.  These probabilities
account for the uncertainties and covariances between the parameters of the radial velocity
fit, the uncertainty in the mass of the primary, assuming a uniform
distribution of $\cos i$, and adopting various priors for the distribution of companion 
mass ratios ${\rm d}N/{\rm d}\log{q}$.
\label{fig:mass}}
\end{figure}

We weight the resulting distribution of $m$ by a prior on the luminosity ratio $l$ and a prior on the mass
ratio $q$.  Specifically, we assume luminosity ratios of $l\ga 0.1$
are excluded by the lack of features due to the companion in the
high-resolution spectra.  We assume a flux ratio relationship of the form $l \propto q^{4.5}$, as is roughly
appropriate for main-sequence stars.  The exponent of 4.5 is derived by fitting the values found in Table 1 of \citet{tor2010} for stars with $0.5 < M_{\odot} < 1.5$.  We note that the precise
value of the exponent for this relationship does not significantly affect our conclusions 
for reasonable values in the range 2.5-6.5.  We therefore
assume the companion is on the main sequence and is not a remnant.
Since the mass ratio distribution for companions in the relevant range
of masses is uncertain, we adopt three different priors that are likely to 
bracket the true distribution (see,
e.g., \citealt{dm91,mazeh92,gre2006}).  Specifically, we assume linear
(${\rm d}N/{\rm d}\log{q} \propto q$), logarithmic (${\rm d}N/{\rm
d}\log{q} \propto \log{q}$), and constant (${\rm d}N/{\rm d}\log{q} =$ constant)
priors.  For the logarithmic and constant priors, a brown dwarf
companion is slightly favored ($\sim 66\%$ and $61\%$, respectively),
whereas for the linear prior, a stellar companion is slightly favored
($57\%$).  Fig.\ \ref{fig:mass} plots the cumulative probability that the mass of the companion is less than a given mass for the three different assumed priors.

\section{Effects of Tides}
Given the proximity of the companion to the host star, tidal interactions could be important in this system.  The tidal effects between solar mass
stars and very low mass stars/brown dwarfs are not
well-studied. Moreover, tidal models themselves are complicated and
uncertain, and observational constraints are few, especially among brown
dwarfs and very late M dwarfs. The situation is further complicated by
the ambiguity of the secondary's mass, which partly determines the
tidal evolution. Nonetheless, we consider the tidal evolution of the
system in this section.

Following \citet{gol1966}, we assume that the tidal response of a body
can be adequately modeled by a single parameter, $Q'$, which is
related to the angle between the lines connecting the centers of the
two bodies, and the center of the deformed body to its tidal bulge.
See \citet{hel2010} or \citet{fer2008} for recent reviews of tidal
theory. Furthermore, we require that as the orbit of the secondary
evolves, the parameter $Q'$ remains constant. If we assume the orbit is circular, it may
be shown that the semi-major axis of the secondary decays as
\begin{equation}
\label{eq:dadt}
\frac{da}{dt} = \frac{9}{2}\sqrt{\frac{G}{M}}\frac{R^5m}{Q'_*}a^{-11/2}
\end{equation}
\citep{gol1966}, where $Q'_*$ is the tidal quality factor of the star divided by two-thirds of its Love number, $M$ is the mass of the primary, $R$ is the radius of the primary and $m$ is the mass of the secondary. Subscript ``*'' refers to the primary.

Equation~\ref{eq:dadt} only applies if the primary's rotation period is
longer than the orbital period. Should the two be equal, the tidal bulges
align, the torques reduce to zero, and the orbital evolution
effectively halts. There still may be evolution due to the obliquity
tide raised on the star, but that evolution is orders of magnitude
slower, and we ignore it here. As the secondary decays, its angular
momentum, $L_{orb} = m\sqrt{GMa}$, decreases and is passed to the
star's rotational angular momentum, $L_* = C_*MR^2\Omega$, where
$\Omega$ is the primary's rotation frequency, and $C_*$ is the
primary's moment of inertia coefficient. We set $C_*$ to
0.1 \citep{mas2008}. Therefore, over a given time, the change in the
primary's rotational frequency is
\begin{equation}
\label{eq:spinup}
\Delta\Omega = -\frac{\Delta L_{orb}}{C_*MR^2_*},
\end{equation}
where $\Delta L_{orb}$ is the change in orbital angular momentum over
the same time. Eqs.~\ref{eq:dadt} and \ref{eq:spinup} can be solved
together to determine when the rotational frequency equals the orbital
frequency (i.e. synchronization) when orbital decay stops.  Here we
ignore the case of the primary's rotation period being shorter than
the orbital period, but in that case, the secondary will spin down the
star, while its orbit expands, also driving the system toward synchronization.

We have examined the tidal evolution of this system in the range
$10^4 \le Q'_* \le 10^{10}$. Most studies find $Q'_*$ values in the
range $10^5$ -- $10^7$ \citep[][but see also \citet{bar2010}]{mat1994,jac2008}, but in reality this
parameter is very poorly constrained. Our chosen range covers all
plausible values. We also vary $i$, the inclination of the secondary's
orbit to the line of sight, for $90^\circ$ -- $1^\circ$ (from edge-on to nearly face-on) and adjust the
mass accordingly. We determine the primary's rotation period via the
measured value of $v \sin I$ and the derived radius of the primary,
$R$. For this analysis, we set the primary's equator in the same
plane as the secondary's orbit (i.e., we assume $i = I$), but this decision does not qualitatively affect our
results.

\begin{figure}
\plotone{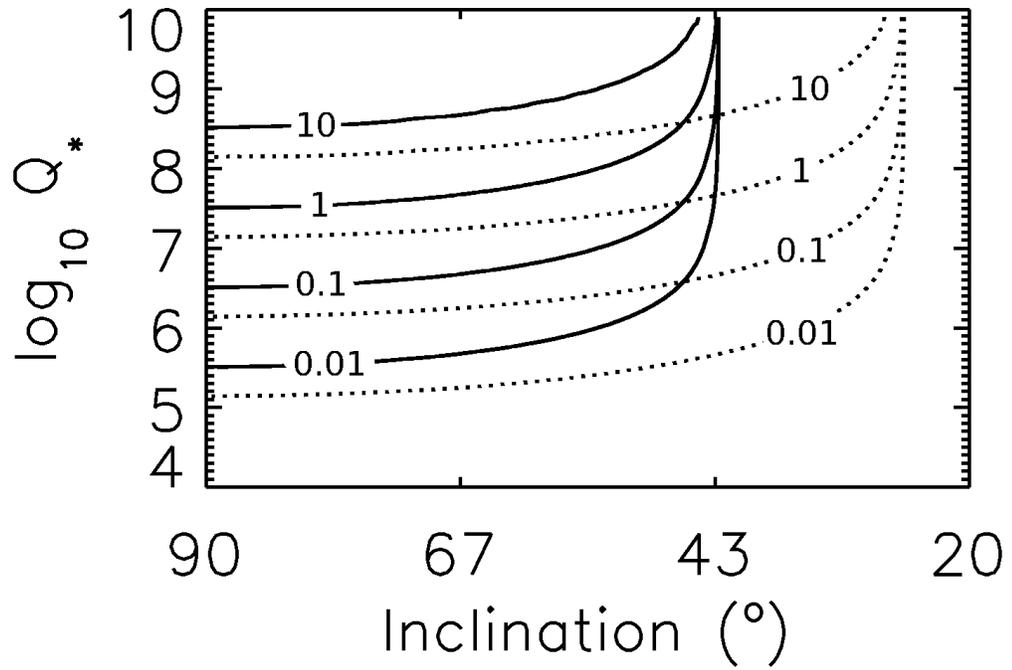}
\caption{Contours of the time required to synchronize the primary's rotational period to the orbital period in Gyr. Solid contours correspond to the best fit, dotted contours are for an unlikely but plausible case which maximizes the synchronization timescale.\label{fig:tides}}
\end{figure}

In Fig.\ \ref{fig:tides} we show the synchronization times from
Eqs.~\ref{eq:dadt}--\ref{eq:spinup} over the parameter space defined
above. We consider two models, the best-fit set of parameters (solid
contours), and one in which $v\sin I = 4$ km s$^{-1}$ (the 3-sigma minimum
value), $M = 1.22$ M$_\odot$, and $R = 1.02$ R$_\odot$ (dotted
contours), i.e. a pathological case, permitted by the observational
uncertainties, with values of $v \sin I $, $M$, and $R$ that
minimize the tidal evolution. First, note the convergence of contour
lines at $i = 43^\circ$ (solid contours) and $i = 25^\circ$ (dotted
contours). These singularities occur because $i$ is small enough that
the system is currently synchronized, i.e. $v \sin I$ equals the
circular velocity of the companion. For $i$ greater than these values the
secondary is spinning up the primary. We find that for a wide range of
$Q'_*$ and $i$ combinations, the secondary quickly spins the primary
up to synchronization. For the best-fit and $Q'_* \lsim 10^7$, the synchronization time is $\lsim 0.1$ Gyr.  The photometry shows no significant variability due to starspots that would indicate a particularly young F star, so it is likely that the age of the system is larger than the synchronization timescale.  We interpret this short timescale as evidence that
the secondary has already spun the primary up to synchronization. For
the pathological case, the time to synchronize could be a factor of a
few larger, but still small compared to the likely age of the system. We do
not show the analogous case for large $v \sin I$, $M$ and $R$,
as that case is already synchronized for all values of $i$.

We conclude that this system is most likely in a double synchronous
state in which the orbit is circular, and both primary and secondary
rotational frequencies are equal to the orbital frequency. We
emphasize that this conclusion is tentative; should more information
regarding $Q'_*$ or the relative orientations of the primary's spin
axis and orbital plane be determined, our analysis will need to be updated.

\section{Conclusion}
We have discovered a brown dwarf candidate around the main sequence F
star TYC 2949-00557-1 during the MARVELS Pilot Project, a wide-area,
multi-object, radial velocity search for planets.  We have
characterized the properties of the host star and the dynamics of the
orbiting companion.  The companion straddles the hydrogen burning
minimum mass with an $m_{\rm{min}} i = 64 ~ M_{J}$, and the orbital period of
$\sim$ 5.7 days places this candidate in the brown dwarf ``orbital
separation desert''.  This desert extends to very low-mass stars
(VLMS) with masses up to $\sim$ 150 ~ $M_{J}$, reflective of the fact
that the formation and dynamical evolution of these objects is
independent of the precise substellar mass limit.  Therefore, TYC
2949-00557-1b is still a ``desert dweller'' for all but the most
extreme face-on inclinations.  Tidal analysis suggests that the system
is in a double synchronous state where both companions have achieved
rotational synchronization in addition to achieving orbital
synchronization.

If the companion's true mass is substellar, this is an example of a
rare class of brown dwarfs at orbital distances comparable to those of
``Hot Jupiters''.  Our photometric observations show no evidence
of variability.  Although we are able to exclude transits for likely companion
radii, the orbital inclination is otherwise unconstrained,
and thus there remains the possibility that this system is a VLMS companion with
a nearly face-on inclination.  Depending on the assumed priors for the
distribution of binary mass ratios, we estimate a probability of
43-66\% that the companion is below the hydrogen burning minimum mass.
The large number of stars surveyed by multi-object, radial velocity
instruments may allow for additional rare systems to be discovered,
including brown dwarfs in the mass or orbital separation deserts.
Such rare systems can provide important constraints to models of the
formation and internal structure of low-mass stars and brown dwarfs,
in addition to elucidating the formation and evolution of extrasolar
planetary systems.

\acknowledgements
We thank the anonymous referee for comments that improved the quality of this paper.  Thanks to Eric Agol for editing of the manuscript draft that greatly improved the quality of the paper.  Funding for the multi-object Doppler instrument was provided by the
W.M. Keck Foundation.  The pilot survey was funded by NSF with grant
AST-0705139, NASA with grant NNX07AP14G and the University of Florida.
S.W.F. is supported through a Florida Space Grant Fellowship.  Keivan Stassun and Joshua Pepper acknowledge funding support from the Vanderbilt Initiative in Data-Intensive Astrophysics (VIDA) from Vanderbilt University, and from NSF Career award AST-0349075.  This research has made use of the SIMBAD database, operated at CDS, Strasbourg, France.  Based on observations with the SDSS 2.5-meter telescope.  Funding for the SDSS and SDSS-II has been provided by the Alfred P. Sloan Foundation, the Participating Institutions, the National Science Foundation, the U.S. Department of Energy, the National Aeronautics and Space Administration, the Japanese Monbukagakusho, the Max Planck Society, and the Higher Education Funding Council for England. The SDSS Web Site is http://www.sdss.org/.  The SDSS is managed by the Astrophysical Research Consortium for the Participating Institutions. The Participating Institutions are the American Museum of Natural History, Astrophysical Institute Potsdam, University of Basel, University of Cambridge, Case Western Reserve University, University of Chicago, Drexel University, Fermilab, the Institute for Advanced Study, the Japan Participation Group, Johns Hopkins University, the Joint Institute for Nuclear Astrophysics, the Kavli Institute for Particle Astrophysics and Cosmology, the Korean Scientist Group, the Chinese Academy of Sciences (LAMOST), Los Alamos National Laboratory, the Max-Planck-Institute for Astronomy (MPIA), the Max-Planck-Institute for Astrophysics (MPA), New Mexico State University, Ohio State University, University of Pittsburgh, University of Portsmouth, Princeton University, the United States Naval Observatory, and the University of Washington.  Based on observations obtained with the Apache Point Observatory 3.5-meter telescope, which is owned and operated by the Astrophysical Research Consortium.  This publication makes use of data products from the Two Micron All Sky Survey, which is a joint project of the University of Massachusetts and the Infrared Processing and Analysis Center/California Institute of Technology, funded by the National Aeronautics and Space Administration and the National Science Foundation.  The Hobby-Eberly Telescope (HET) is a joint project of the University of Texas at Austin, Pennsylvania State University, Stanford University, Ludwig-Maximillians-Universit\"{a}t M\"{u}nchen, and Georg-August-Universit\"{a}t G\"{o}ttingen. The HET is named in honor of its principal benefactors, William P. Hobby and Robert E. Eberly.  The Center for Exoplanets and Habitable Worlds is supported by the Pennsylvania State University, the Eberly College of Science, and the Pennsylvania Space Grant Consortium.

\begin{deluxetable}{rrr}
\tabletypesize{\scriptsize}
\tablecaption{MPP RV Observations.\label{aporvs}}
\tablewidth{0pt}
\tablehead{
\colhead{BJD$_{\rm{TDB}}$} & \colhead{RV} & \colhead{$\sigma_{RV}$\tablenotemark{a}}\\
\colhead{~} & \colhead{(m s$^{-1}$)} & \colhead{(m s$^{-1}$)}
}
\startdata
2454101.69079 & 13339 & 94 \\
2454105.75520 & 19819 & 98 \\
2454106.70228 & 14981 & 96 \\
2454128.62211 & 19018 & 143 \\
2454128.86491 & 17170 & 91 \\
2454130.83623 & 13299 & 94 \\
2454136.64529 & 15043 & 85 \\
2454136.85894 & 15854 & 139 \\
2454163.72245 & 14882 & 91 \\
2454188.69749 & 21003 & 136 \\
2454191.68122 & 15421 & 97 \\
2454194.69216 & 21197 & 115 \\
2454195.68446 & 24453 & 102 \\
2454217.61115 & 22758 & 79 \\
\enddata
\tablenotetext{a}{Errors are not scaled to account for systematics.}
\end{deluxetable}

\begin{deluxetable}{rrr}
\tabletypesize{\scriptsize}
\tablecaption{KPNO/ET RV Observations.\label{kpnorvs}}
\tablewidth{0pt}
\tablehead{
\colhead{BJD$_{\rm{TDB}}$} & \colhead{RV} & \colhead{$\sigma_{RV}$\tablenotemark{a}}\\
\colhead{~} & \colhead{(m s$^{-1}$)} & \colhead{(m s$^{-1}$)}
}
\startdata
2454749.97070 &   266 & 58 \\
2454751.88336 &  1316 & 55 \\
2454751.94579 &  1985 & 50 \\
2454752.89657 &  7580 & 54 \\
2454752.95821 &  8026 & 58 \\
2454753.90161 &  9651 & 77 \\
2454753.96355 &  9548 & 65 \\
2454754.90388 &  4739 & 53 \\
2454754.96580 &  4502 & 52 \\
2454755.91284 &  -694 & 47 \\
2454755.97519 &  -996 & 52 \\
\enddata
\tablenotetext{a}{Errors are not scaled to account for systematics.}
\end{deluxetable}

\begin{deluxetable}{rrr}
\tabletypesize{\scriptsize}
\tablecaption{HET/HRS RV Observations.\label{hetrvs}}
\tablewidth{0pt}
\tablehead{
\colhead{BJD$_{\rm{TDB}}$} & \colhead{RV} & \colhead{$\sigma_{RV}$\tablenotemark{a}}\\
\colhead{~} & \colhead{(m s$^{-1}$)} & \colhead{(m s$^{-1}$)}
}
\startdata
2454807.74076 & 0     & 69 \\
2454808.98051 & 4674  & 67 \\
2454825.69507 & 2469  & 107 \\
2454829.70100 & 2482  & 66 \\
2454881.75734 & -23   & 286 \\
2454882.78369 & 3178  & 51 \\
2454883.76208 & 9451  & 63 \\
2454887.75592 & 165   & 48 \\
2454889.74685 & 10881 & 51 \\
2454890.75523 & 11513 & 47 \\
\enddata
\tablenotetext{a}{Errors are not scaled to account for systematics.}
\end{deluxetable}

\begin{deluxetable}{lllll}
\renewcommand{\arraystretch}{1.2}
\tabletypesize{\scriptsize}
\tablecaption{Best-fit dynamical properties of TYC 2949-00557-1.\label{finalresults:orbit}}
\tablewidth{0pt}
\tablehead{
\colhead{Parameter} & \colhead{Value} & \colhead{Uncertainty} & \colhead{Value} & \colhead{Uncertainty}
}
\startdata
~ & \multicolumn{2}{c}{ECCENTRIC CASE} & \multicolumn{2}{c}{CIRCULAR CASE} \\
\hline
Period (days) & 5.69459 & 0.00029 & 5.69449 & 0.00023 \\
K (km s$^{-1}$) & 6.109 & 0.014  & 6.113 & 0.009 \\
$T_{c}$ (BJD$_{\rm{TDB}}$-2454000) & 868.9878 & 0.0042 & 868.9877 & 0.0016 \\
$e\cos{\omega}$ & 0.0000 & 0.0015 & 0. & - \\
$e\sin{\omega}$ & -0.0005 & $^{+0.0011} _{-0.0019}$ & 0. & - \\
$e$ & 0.0017 & $^{+0.0019} _{-0.0017}$ & 0. & - \\
$\omega$ (rad) & 4.69 & 1.58 & $\frac{\pi}{2}$ & - \\
$m_{\rm{min}} i$ (M$_{J}$) & 64.3 & 3.0 & 64.3 & 3.0 \\
Systemic velocity $\gamma_{0}$ (km s$^{-1}$) & 18.67 & 0.26 & 18.68 & 0.24 \\
KPNO offset ($\gamma_{0}-\gamma_{kpno}$, km s$^{-1}$) & 14.89 & 0.26 & 14.90 & 0.25 \\
HET offset ($\gamma_{0}-\gamma_{het}$, km s$^{-1}$) & 12.59 & 0.26 & 12.61 & 0.24 \\
MPP slope $\dot\gamma_{mpp}$ (km s$^{-1}$ day$^{-1}$) & -0.0044 & 0.0065 & -0.0037 & 0.0065 \\
KPNO slope $\dot\gamma_{kpno}$ (km s$^{-1}$ day$^{-1}$) & 0.016 & 0.034 & 0.013 & 0.033 \\
HET slope $\dot\gamma_{het}$ (km s$^{-1}$ day$^{-1}$) & -0.0010 & 0.00035 & -0.0011 & 0.00024 \\
Total $\sigma_{RV}$ Scale Factor (MPP)  & 8.64 & - & 8.49 & - \\
Total $\sigma_{RV}$ Scale Factor (KPNO) & 3.43 & - & 3.37 & - \\
Total $\sigma_{RV}$ Scale Factor (HET)  & 0.39 & - & 0.30 & - \\
Combined $\chi^2/{\rm dof}$ & 1.69 & - & 1.55 & - \\
\enddata
\end{deluxetable}

\begin{deluxetable}{lll}
\renewcommand{\arraystretch}{1.2}
\tabletypesize{\scriptsize}
\tablecaption{Stellar host properties of TYC 2949-00557-1.\tablenotemark{a}\label{finalresults:mags}}
\tablewidth{0pt}
\tablehead{
\colhead{Parameter} & \colhead{Value} & \colhead{Uncertainty} \\
}
\startdata
$\alpha$ (J2000)\tablenotemark{b} & 101.921152 (deg)  & 06:47:41.076 (HH:MM:SS)  \\
$\delta$ (J2000)\tablenotemark{b} & 42.009332~ (deg)  & 42:00:33.60~ (DD:MM:SS)  \\
$B$ & 12.846 & 0.023 \\
$V$ & 12.142 & 0.031  \\
$R_{c}$ & 11.750 & 0.039 \\
$I_{c}$ & 11.391 & 0.043 \\
$J_{2MASS}$ & 10.820 & 0.022 \\
$H_{2MASS}$ & 10.474 & 0.021 \\
$K_{2MASS}$ & 10.421 & 0.018 \\
$T_{\rm eff}$ (K) & 6135.  & 40. \\
$\log{(g ~ {\rm [cm s^{-1}]})}$ & 4.4  & 0.1 \\
$[Fe/H]$ & 0.32  & 0.01 \\
$v \sin I$ (km s$^{-1}$) & 7. & 1. \\
$M_{primary}$ ($M_{\odot}$) & 1.25 & 0.09 \\
$R_{primary}$ ($R_{\odot}$) & 1.15 & 0.15 \\
$d$ (pc) & 413 & $^{+66} _{-57}$ \\
\enddata
\tablenotetext{a}{$BVRI$ magnitudes are unweighted averages from Table \ref{multimagtable}.}
\tablenotetext{b}{Coordinates taken from the Tycho 2 Catalog \citep{hog2000}}
\end{deluxetable}

\begin{deluxetable}{lcc}
\tabletypesize{\scriptsize}
\tablecaption{Measured photometry for TYC 2949-00557-1 from Hereford Arizona Observatory.\label{multimagtable}}
\tablewidth{0pt}
\tablehead{
\colhead{Band\tablenotemark{a}} & \colhead{Flux} & \colhead{Uncertainty}\\
\colhead{~} & \colhead{(mags)} & \colhead{(mags)}
}
\startdata
$B$ & 12.839 & 0.023 \\
$V$ & 12.135 & 0.031 \\
$Rc$ & 11.750 & 0.039 \\
$Ic$ & 11.390 & 0.043 \\
$g^\prime$ & 12.450 & 0.020 \\
$r^\prime$ & 11.937 & 0.016 \\
$i^\prime$ & 11.811 & 0.028 \\
$B_{Sloan}$ & 12.853 & 0.023 \\
$V_{Sloan}$ & 12.150 & 0.024 \\
$Rc_{Sloan}$ & 11.749 & 0.029 \\
$Ic_{Sloan}$ & 11.392 & 0.043 \\
\enddata
\tablenotetext{a}{Measurements labeled with $_{Sloan}$ are converted from the Sloan filter observations using \citet{smi2002} transformation equations.}
\end{deluxetable}
\clearpage


\end{document}